%% file: ANA-STDM-2017-29-PAPER.tex
\pdfoutput=1
 
\newcommand*{\ATLASLATEXPATH}{}
\documentclass[cernpreprint,texlive=2016,txfonts,UKenglish]{\ATLASLATEXPATH atlasdoc}
 
\usepackage{\ATLASLATEXPATH atlaspackage}
\usepackage{\ATLASLATEXPATH atlasphysics}
 
\newcommand{\AtlasCoordFootnote}{ATLAS uses a right-handed coordinate system
with its origin at the nominal interaction point (IP) in the centre of the
detector and the $z$-axis along the beam pipe. The $x$-axis points from
the IP to the centre of the LHC ring, and the $y$-axis points upwards.
Cylindrical coordinates $(r,\phi)$ are used in the transverse plane, $\phi$
being the azimuthal angle around the $z$-axis. The transverse energy is
defined as $E_{\mathrm{T}}=E\sin\theta$, where $E$ is the energy and
$\theta$ is the polar angle. The pseudorapidity is defined as
$\eta=-\ln\tan(\theta/2)$ and the angular distance is measured in units
of $\Delta R\equiv\sqrt{(\Delta\eta)^{2}+(\Delta\phi)^{2}}$.
}
 
\graphicspath{{logos/}}
 
\usepackage{\ATLASLATEXPATH atlasbiblatex}
 
\def\etaphi{\eta\text{--}\phi}
 
\def\pb1{pb$^{-1}$}
\def\fb1{fb$^{-1}$}
 
\def\qq{q\bar q}
 
\def\dgen{\frac{{\mathrm{d}}\sigma}{{\mathrm{d}}\etg}}
\def\dgenn{{\mathrm{d}}\sigma/{\mathrm{d}}\etg}
 
\def\ppg{pp\rightarrow\gamma\ +\ {\mathrm{X}}}

\def\mz{m_{\mathrm{Z}}}
 
\def\as{\alpha_{\mathrm{s}}}
\def\asz{\alpha_{\mathrm{s}}(\mz)}
 
\def\muR{\mu_{\mathrm{R}}}
\def\muF{\mu_{\mathrm{F}}}
\def\muf{\mu_{\mathrm{f}}}

\def\etg{E_{\mathrm{T}}^{\gamma}}
\def\etag{\eta^{\gamma}}
\def\pt{p_{\mathrm{T}}}
 
\def\etiso{E_{\mathrm{T}}^{\mathrm{iso}}}
\def\etisoc{E_{\mathrm{T,cut}}^{\mathrm{iso}}}
\def\etisocut{$4.2\times 10^{-3}\times\etg+4.8\ {\mathrm{\GeV}}$}
 
\def\sher{{\textsc{Sherpa}}}
\def\pyt{{\textsc{Pythia}}}
\def\jetp{{\textsc{Jetphox}}}
\def\nnlojet{{\textsc{Nnlojet}}}
 
\AtlasTitle{Measurement of the inclusive isolated-photon cross section
in {\boldmath $pp$} collisions at {\boldmath $\sqrt s=13$~\TeV} using
$36$~\fb1\ of ATLAS data}
 
\author{The ATLAS Collaboration}
 
\AtlasJournalRef{JHEP 10 (2019) 203}
\AtlasDOI{10.1007/JHEP10(2019)203}
\PreprintIdNumber{CERN-EP-2019-136}
 
\AtlasAbstract{
The differential cross section for isolated-photon production in $pp$
collisions is measured at a centre-of-mass energy of $13$~\TeV\ with
the ATLAS detector at the LHC using an integrated luminosity of
$36.1$~\fb1. The differential cross section is presented as a
function of the photon transverse energy in different regions of
photon pseudorapidity. The differential cross section as a function of
the absolute value of the photon pseudorapidity is also presented in
different regions of photon transverse energy. Next-to-leading-order
QCD calculations from \jetp\ and \sher\ as well as
next-to-next-to-leading-order QCD calculations from \nnlojet\ are
compared with the measurement, using several parameterisations of the
proton parton distribution functions. The predictions provide a good
description of the data within the experimental and theoretical
uncertainties.
}
 
\hypersetup{pdftitle={ATLAS document},pdfauthor={The ATLAS Collaboration}}
 
\begin{document}
 
\maketitle
 
\tableofcontents
 
\section{Introduction}
\label{intro}
Prompt photons with large transverse momenta constitute colourless
probes of the hard interaction and their production in proton--proton
collisions, $\ppg$, provides a testing ground for perturbative QCD
(pQCD). Prompt photons are defined as those that are not secondaries
from hadron decays. Prompt-photon production is understood to proceed
via two processes: the one in which the photon arises directly from
the hard interaction (the direct-photon process) and the one in which
the photon is emitted in the fragmentation of a high transverse
momentum ($\pt$) parton (the fragmentation-photon
process)~\cite{pr:d76:034003,pr:d79:114024}.
 
In proton--proton collisions, due to the abundance of photons from
neutral-hadron decays and the contribution from the fragmentation
process, prompt-photon production is studied by requiring the photons
to be isolated. Measurements of inclusive isolated-photon production
in $pp$ collisions at centre-of-mass energies ($\!\!\sqrt s$) of $7$,
$8$ and $13$~\TeV\ are available from the
ATLAS~\cite{pr:d89:052004,jhep:1608:005,pl:b770:473} and
CMS~\cite{pr:d84:052011,epj:c79:20} collaborations.
 
At the LHC the dominant contribution to prompt-photon production
arises from the $qg\rightarrow q\gamma$ process. As a consequence, the
production is sensitive to the gluon density in the
proton~\cite{np:b860:311,epl:101:61002,pr:742:1} even at leading order
(LO) in pQCD. A recent study~\cite{epj:c78:470} has included the ATLAS
measurement of inclusive isolated-photon production at
8~\TeV~\cite{jhep:1608:005} in a next-to-next-to-leading-order (NNLO)
QCD fit to obtain the parton distribution functions (PDF) of the
proton. The inclusion of the ATLAS data leads to a reduction in the
gluon density uncertainties~\cite{epj:c78:470}.
 
Measurements of inclusive prompt-photon production also provide
benchmarks to use in investigating novel approaches to parton
radiation~\cite{pr:d94:034020}, next-to-leading-order (NLO) QCD
corrections with matched parton showers~\cite{jhep:1803:081}, the
relevance of threshold logarithms in QCD and electroweak
corrections~\cite{jhep:1609:005} and NNLO QCD
corrections~\cite{prl:118:222001,1904.01044,1905.08577}. More
specifically, the production of prompt photons, which is less
sensitive to hadronisation effects than that of jets, can be used to
pursue an alternative QCD description based on the $k_\textrm{T}$
factorisation approach combining off-shell amplitudes and
transverse-momentum-dependent parton
densities~\cite{pr:d94:034020}. The prompt-photon data allow the
investigation of the fragmentation contribution, which can be done
either via fragmentation functions or through the inclusion of parton
showers~\cite{jhep:1803:081}. Electroweak corrections are likely to
play an important role at the \TeV\ scale and, thereby, measurements
at photon transverse energies in that region would help to unveil such
phenomena~\cite{jhep:1609:005}. The prompt-photon measurements at the
LHC  are characterised by small uncertainties that demand precise
theoretical predictions, e.g.\ those from NNLO QCD
calculations~\cite{prl:118:222001,1904.01044,1905.08577}, in order to
fully exploit these data. Measurements involving isolated photons have
also been used to constrain the contributions from new light scalar
particles decaying into a photon pair~\cite{1803.10379}.
 
This paper presents a measurement of isolated prompt-photon production
in $pp$ collisions at $\sqrt s=13$~\TeV\ with the ATLAS detector at
the LHC using an integrated luminosity of $36.1$~\fb1. The
differential cross section as a function of the photon transverse
energy\footnote{\AtlasCoordFootnote} ($\etg$) is measured in different
regions of the photon pseudorapidity ($\etag$) for $\etg>125$~\GeV\
and $|\etag|<2.37$, excluding the region $1.37<|\etag|<1.56$. In
addition, the double-differential cross section as a function of
$|\etag|$ in different regions of $\etg$ is also presented. The
results are based on a data sample with a more than ten-fold increase
in statistics relative to the previous study~\cite{pl:b770:473}. The
measurement presented here is found to be consistent with the previous
one in the overlapping kinematic regions. This increase in statistics
allows improvements in the calibration of the photon energy and
reductions in the experimental systematic uncertainties affecting the
cross-section measurement, as well as an extension of the coverage in
$\etg$ to higher values than previously measured. In this analysis,
the region where the measurement is limited by systematic
uncertainties is extended to $\etg\sim 1$~\TeV, beyond what was
achieved in the previous measurement. The NLO QCD predictions of
\jetp~\cite{jhep:0205:028,pr:d73:094007} and
\sher~\cite{jhep:0902:007} based on several parameterisations of the
PDFs are compared with the measurement.  The NNLO QCD prediction of
\nnlojet~\cite{1904.01044}, which has significantly reduced
uncertainties due to fewer missing higher-order terms, is also
confronted with the data.
 
\section{ATLAS detector}
\label{detector}
The ATLAS detector~\cite{PERF-2007-01,ATLAS-TDR-19,Abbott:2018ikt} is
a multipurpose  detector with a forward--backward symmetric
cylindrical geometry. It consists of an inner tracking detector
surrounded by a thin superconducting solenoid, electromagnetic and
hadronic calorimeters, and a muon spectrometer incorporating three
large superconducting toroid magnets. The inner-detector system is
immersed in a $2$~T axial magnetic field and provides charged-particle
tracking in the range $|\eta|<2.5$. The high-granularity silicon pixel
detector is closest to the interaction region and provides four
measurements per track. The pixel detector is followed by the silicon
microstrip tracker, which typically provides four three-dimensional
space point measurements per track. These silicon detectors are
complemented by the transition radiation tracker, which enables
radially extended track reconstruction up to $|\eta|=2$. The
calorimeter system covers the range $|\eta|<4.9$. Within the region
$|\eta|<3.2$, electromagnetic (EM) calorimetry is provided by barrel
and endcap high-granularity lead/liquid-argon (LAr) calorimeters, with
an additional thin LAr presampler covering $|\eta|<1.8$ to correct for
energy loss in material upstream of the calorimeters; for
$|\eta|<2.5$, the EM calorimeter is divided into three layers in
depth. Hadronic calorimetry is provided by a steel/scintillator-tile
calorimeter, segmented into three barrel structures within
$|\eta|<1.7$, and two copper/LAr hadronic endcap calorimeters, which
cover the region $1.5<|\eta|<3.2$. The solid-angle coverage is
completed out to $|\eta|=4.9$ with forward copper/LAr and tungsten/LAr
calorimeter modules, which are optimised for EM and hadronic
measurements, respectively. Events are selected using a first-level
trigger implemented in custom electronics, which reduces the maximum
bunch crossing rate of $40$~MHz to a design value of $100$~kHz using a
subset of detector information. Software algorithms with access to the
full detector information are then used in the high-level trigger to
yield a recorded event rate of about $1$~kHz~\cite{epj:c77:317}.
 
\section{Data sample and Monte Carlo simulations}
\label{datmc}
The data used in this analysis were collected with the ATLAS detector
during the proton--proton collision running periods of 2015 and 2016,
when the LHC operated at a centre-of-mass energy of $\sqrt
s=13$~\TeV. Only events taken during stable beam conditions and
satisfying detector- and data-quality requirements, which include the
calorimeters and inner tracking detectors being in nominal operation,
are considered. In the collected data sample the average number of
$pp$ interactions per bunch crossing is $24$, while the total
integrated luminosity is {\mbox{$36.1\pm 0.8$~\fb1}} out of which
{\mbox{$3.22\pm 0.07$~\fb1}} corresponds to the 2015 running
period. The uncertainty in the combined 2015--2016 integrated
luminosity is  $2.1\%$~\cite{ATLAS-CONF-2019-021}, obtained using the
LUCID-2 detector~\cite{LUCID2} for the primary luminosity
measurements.
 
The simulated events were produced using MC event generators and were
passed through the {\textsc{Geant4}}-based~\cite{nim:a506:250} ATLAS
detector and trigger simulation programs~\cite{epj:c70:823}. They are
reconstructed and analysed with the same programs chain as the data.
 
{\textbf{Simulated signal samples.}}
Samples of prompt-photon events were generated using the programs
\pyt\ 8.186~\cite{cpc:178:852} and \sher~2.1.1~\cite{jhep:0902:007} to
study the characteristics of signal events. In both cases, the event
generation was performed using tree-level matrix elements, with the
inclusion of initial- and final-state parton showers. Fragmentation
into hadrons is described by the Lund string model~\cite{prep:97:31}
in the case of \pyt\ and by a modified version of the cluster
model~\cite{epj:c36:381} in the case of \sher. The proton's structure
was parameterised by the LO NNPDF2.3~\cite{np:b867:244} PDF set  for
\pyt\ and by the NLO CT10~\cite{pr:d82:074024} PDF set for \sher. All
samples include a simulation of the underlying event (UE), with
parameter values set according to the ATLAS 2014 tune series (A14
tune) for \pyt~\cite{ATL-PHYS-PUB-2014-021} or to the tune developed
by the authors of \sher\ for use in conjunction with the NLO CT10 PDF
set. The \pyt\ simulation of the signal includes LO photon-plus-jet
events from direct processes (the hard subprocesses $qg\rightarrow
q\gamma$ and $\qq\rightarrow g\gamma$, called the `hard' component)
and photon bremsstrahlung in LO QCD dijet events (called the
`bremsstrahlung' component). The bremsstrahlung component is modelled
by final-state QED radiation arising from calculations of all
$2\rightarrow 2$ QCD processes. The \sher\ samples were generated with
LO matrix elements for photon-plus-jet final states with up to three
additional partons ($2\rightarrow n$ processes with $n$ from $2$ to
$5$); the matrix elements were merged with the \sher\ parton shower
using the ME+PS@LO prescription~\cite{jhep:0905:053}. The
bremsstrahlung component is accounted for in \sher\ through the matrix
elements of $2\rightarrow n$ processes with $n\geq 3$. In the
generation of the \sher\ samples, a requirement on the photon
isolation at the matrix-element level is imposed using the criterion
defined in Ref.~\cite{pl:b429:369}. This criterion, commonly called
Frixione's criterion, requires the total transverse energy inside a
cone of size ${\mathfrak r}$ in the $\etaphi$ plane around the
generated final-state photon, excluding the photon itself, to be below
a certain threshold,
$E_{\mathrm{T}}^{\mathrm{max}}({\mathfrak r})
=\epsilon\etg ((1-\cos{{{\mathfrak r}}})/(1-\cos{{\cal R}}))^{n}$,
for all ${\mathfrak r} < {\cal R}$, where ${\cal R}$
is the maximal cone size, $n$ is the power and $\epsilon$ is a
constant such that $\epsilon\etg$ represents the threshold for
${\mathfrak r}={\cal R}$. The parameters for the threshold are chosen
to be ${\cal R}=0.3$, $n=2$ and $\epsilon=0.025$. The criterion is
applied to avoid divergencies in the matrix elements when the photon
is collinear with a parton.
 
{\textbf{Simulated background samples.}}
The main background to isolated-photon events arises from jets
misidentified as photons. This background is subtracted using a
data-driven technique, which is described in Section~\ref{bacsub};
thus, no MC sample is used to simulate this background. The background
from electrons or positrons misidentified as photons is evaluated
using MC samples generated with the program
\sher~2.2.1~\cite{jhep:0902:007,jhep:12:039,jhep:0202:044,prl:108:111601,jhep:0803:038,jhep:1304:027}.
The $pp \rightarrow Z/\gamma^*\rightarrow e^+e^- + {\mathrm{X}}$ and
$pp \rightarrow  W\rightarrow e\nu + {\mathrm{X}}$ processes were
generated with matrix elements calculated for up to two additional
partons at NLO and up to four partons at LO. The NNLO NNPDF3.0 PDF
set~\cite{jhep:1504:040} was used in conjunction with a dedicated set
of parton-shower-generator parameter values (tune) developed by the
\sher\ authors.
 
{\textbf{Simulation of pile-up.}}
Pile-up from additional $pp$ collisions in the same and neighbouring
bunch crossings was simulated by overlaying each MC event with a
variable number of simulated inelastic $pp$ collisions generated using
\pyt~8.186 with the ATLAS set of tuned parameters for minimum-bias
events (A2 tune)~\cite{ATL-PHYS-PUB-2012-003} and the MSTW2008LO PDF
set~\cite{epj:c63:189}. The MC events are weighted (`pile-up
reweighting') to reproduce the distribution of the average number of
interactions per bunch crossing observed in the data.
 
\section{Event selection}
\label{evesel}
Events were recorded using a single-photon trigger with a transverse
energy threshold of $120$~\GeV\ ($140$~\GeV) for the 2015 (2015 and
2016) data-taking period\footnote{The single-photon trigger with the
threshold of 120~\GeV\ was kept unprescaled during 2015 whereas it
had to be prescaled during 2016 due to the increase in instantaneous
luminosity.} and `loose' photon identification
requirements~\cite{epj:c77:317,epj:c79:205}.\footnote{The
discriminating variables used for `loose' and `tight' photon
identification can be found in Table~1 of Ref.~\cite{epj:c79:205}.}
Events are required to contain at least one reconstructed
proton--proton interaction vertex. The vertex with the highest sum of
the $\pt^2$ of the associated tracks is selected as the primary vertex.
 
Photon candidates are reconstructed from clusters of energy deposited
in the EM calorimeter and classified~\cite{epj:c79:205} as unconverted
photon candidates (clusters without a matching track or without a
matching reconstructed conversion vertex in the inner detector) or
converted photon candidates (clusters with a matching reconstructed
conversion vertex or a matching track consistent with originating from
a photon conversion).  The main background in the prompt-photon
production measurement comes from an energetic $\pi^0$ or $\eta$ meson
which is misidentified as a photon because it decays into an almost
collinear photon pair. Such energetic $\pi^0$ or $\eta$ mesons are
produced copiously inside jets. This background is reduced by the
photon identification criteria and by requiring the photon candidate
to be isolated. Photon candidates are identified by using variables
that characterise the lateral and longitudinal electromagnetic shower
development in the EM calorimeter and the energy fraction leaking into
the hadronic calorimeter. Tight requirements are imposed on the shower
shapes in the second layer and in the finely segmented first layer of
the EM calorimeter as well as on the energy deposited in the hadronic
calorimeter~\cite{epj:c79:205}. These requirements are optimised
separately for unconverted and converted photon candidates and ensure
the compatibility of the measured shower profile with that originating
from a single photon impacting the calorimeter. Small differences in
the average values of the shower-shape variables between data and
simulation are observed and corrected for in simulated events  prior
to the application of the photon identification criteria. Three
data-driven methods based on radiative $Z$ decays, electron
extrapolation and an inclusive sample of photon
candidates~\cite{epj:c79:205} are used to measure the efficiency of
the tight identification criteria. The results of the three methods
agree in the overlapping kinematic regions and the measured
efficiencies are above $90\%$ ($95\%$) for unconverted (converted)
photon candidates with $\etg >125$~\GeV. Efficiency scale factors are
evaluated as the ratios of the measured efficiencies to the
efficiencies obtained in simulation and are applied to photon
candidates in simulated events. The resulting efficiency scale factors
are compatible with unity and have uncertainties in the range
$1\%$--$3\%$, depending on $\etg$ and $\etag$.
 
The photon isolation requirement is based on the amount of transverse
energy ($\etiso$) inside a cone of size $\Delta R=0.4$ around the
photon candidate, excluding an area of size
$\Delta\eta\times\Delta\phi=0.125\times0.175$ centred on the
barycentre of the photon cluster. The measured value of $\etiso$ is
computed from topological clusters of calorimeter
cells~\cite{epj:c77:490} and is corrected for the expected leakage of
the photon energy into the isolation cone as well as for the
contributions from the UE and
pile-up~\cite{pr:d83:052005,pl:b706:150}. The correction to $\etiso$
from the combined effect of the UE and pile-up is evaluated using the
jet-area method~\cite{jhep:0804:005,jhep:1004:065} on an
event-by-event basis and is typically $3.5$~\GeV\ ($1.3$~\GeV) in the
region $|\etag|<1.37$ ($1.56<|\etag|<2.37$).  In simulated events,
$\etiso$ is adjusted so that the peak position in the $\etiso$
distribution coincides in data and simulation. After the corrections,
$\etiso$ is required to be less than
$\etisoc\equiv$~\etisocut~\cite{jhep:1608:005}.
 
The calibration of the photon energy in the calorimeter accounts for
upstream energy loss as well as lateral and longitudinal leakages. The
procedure used for photon energy calibration in
Run~1~\cite{epj:c74:3071} is employed here, optimised for the detector
configuration in Run~2~\cite{JINST14:P03017}. The energy of the
cluster of calorimeter cells associated with the photon candidate is
corrected using a combination of simulation-based and data-driven
calibration factors determined from $Z\rightarrow e^+e^-$ events
collected during the 2015 and 2016 data-taking periods. The
calibration is performed separately for converted and unconverted
candidates. The uncertainty in the photon energy scale at $\etg =
125$~\GeV\ varies in the range $0.4\%$--$3.1\%$ ($0.4\%$--$2.7\%$) for
unconverted (converted) candidates depending on $|\etag|$.
 
Events with at least one photon candidate with calibrated $\etg$ above
$125$~\GeV\ and $|\etag|<2.37$ are selected. Candidates in the region
$1.37<|\etag|<1.56$, which includes the transition region between the
barrel and endcap calorimeters, are not considered. Photon candidates
with $125< \etg <150$~\GeV\ ($\etg> 150 $~\GeV) are selected from the
2015 (2015 and 2016) data-taking period.
 
If an event contains more than one photon candidate satisfying the
selection criteria above, only the highest-$\etg$ (leading) photon is
considered for further study. The total number of selected events with
a photon candidate with $\etg>150$~\GeV\ ($125<\etg <150$~\GeV) is
$7\,012\,797$ ($650\,410$). The kinematic requirements and the number
of selected events in data in each $|\etag|$ region are summarised in
Table~\ref{tablezero}.  Each region in $|\etag|$ is divided into $16$
bins of $\etg$ starting at $\etg=125$~\GeV\ and ending at
$2500$~\GeV. The binning is given by the following array of values of
$\etg$ (in units of \GeV): $125$, $150$, $175$, $200$, $250$, $300$,
$350$, $400$, $470$, $550$, $650$, $750$, $900$, $1100$, $1500$,
$2000$ and $2500$. Some of the high-$\etg$ bins are not measured
depending on the $|\etag|$ region.
 
\begin{table}[h]
\caption{
Definition of the phase-space region for the measurement and
predictions. The (next-to) last row indicates the number of data
events selected in each $|\etag|$ region for $\etg>150$~\GeV\
($125<\etg <150$~\GeV). For $125<\etg <150$~\GeV\ only data from
the 2015 running period are used while for $\etg>150$~\GeV\ data
from the 2015 and 2016 running periods are used.
}
\setlength{\unitlength}{1.0cm}
\begin{picture} (18.0,5.5)
\put (0.0,2.0){\scalebox{1.0}{\centerline{
\begin{tabular}{|l|c|c|c|c|} \hline
& \multicolumn{4}{c|}{Phase-space region} \\ \hline
& \multicolumn{4}{c|}{ }
\\[-5pt]
Requirement on $\etg$ & \multicolumn{4}{c|}{ $\etg> 125$~\GeV\ }
\\[5pt] \hline
& \multicolumn{4}{c|}{ }
\\[-5pt]
Isolation requirement & \multicolumn{4}{c|}{  $\etiso<$~\etisocut }
\\[5pt] \hline
&  &  & &
\\[-5pt]
Requirement on $|\etag|$ & $|\etag|<0.6$ & $0.6<|\etag|<1.37$ &
$1.56<|\etag|<1.81$ & $1.81<|\etag|<2.37$
\\[5pt] \hline
Number of events with & &  &  &  \\
$125<\etg <150$~\GeV\ & $182\,754$ & $248\,538$ & $74\,405$ & $144\,713$ \\[5pt]  \hline
Number of events with & &  &  &  \\
$\etg>150$~\GeV\ & $2\,030\,144$ & $2\,696\,077$ & $814\,623$ & $1\,471\,953$ \\[5pt]  \hline
\end{tabular}
}}}
\end{picture}
\label{tablezero}
\end{table}
 
\section{Background evaluation and signal extraction}
\label{bacsub}
After the event selection described above, there is a residual
background contribution from jets misidentified as photons. This
background is evaluated bin-by-bin using a data-driven
technique~\cite{jhep:1608:005,pl:b770:473} and subtracted to obtain
the signal yield. For this purpose, a two-dimensional sideband method
is applied, based on a plane formed by the variable $\etiso$ and  the
photon identification criteria.  A photon candidate that fulfils the
tight identification criteria is classified as `tight'; the
requirements are described in Ref.~\cite{epj:c79:205}. A photon
candidate is classified as `non-tight' if it satisfies a modified set
of requirements, but fails the tight identification. The modified set
of requirements is built from the full list of tight requirements by
removing four\footnote{The four variables are $w_{s\ 3}$,
$f_{\mathrm{side}}$, $\Delta E_{\mathrm{s}}$ and
$E_{\mathrm{ratio}}$~\cite{epj:c79:205}. These variables make use of
the first layer of the EM calorimeter; this layer is segmented into
high-granularity strips in the $\eta$ direction. The variable $w_{s\ 3}$
is the lateral shower width calculated from three strips around the
strip with maximum energy deposit. The variable $f_{\mathrm{side}}$
is the energy outside the core of the three central strips but
within seven strips divided by the energy within the three central
strips. The variable $\Delta E_{\mathrm{s}}$ is the difference
between the energy associated with the second maximum in the strip
layer and the energy reconstructed in the strip with the minimum
value found between the first and second maxima. The variable
$E_{\mathrm{ratio}}$ is the ratio of the energy difference between
the maximum energy deposit and the energy deposit in the secondary
maximum in the cluster to the sum of these energies.} of the
selections associated with the shower-shape variables computed from
the energy deposits in the first layer of the EM calorimeter. The
variables that are removed from the list of tight requirements are
those that are least correlated with $\etiso$. Four regions are
defined in the $\etiso$--tightness plane: a signal region ($A$)
containing isolated ($\etiso<\etisoc$) tight photons; a background
control region $B$ consisting of non-isolated ($\etisoc+2\
{\mathrm{\GeV}}<\etiso<50$~\GeV) tight photons; a background control
region $C$ containing isolated non-tight photons; and a background
control region $D$ consisting of non-isolated non-tight photons.
 
The signal yield $N_A^{\mathrm{sig}}$ in region $A$ is extracted by
solving the equation
 
\begin{equation}
N_A^{\mathrm{sig}}=N_A-R^{\mathrm{bg}}\cdot(N_B-f_B N_A^{\mathrm{sig}})\cdot\frac{(N_C-f_C N_A^{\mathrm{sig}})}{(N_D-f_D N_A^{\mathrm{sig}})},
\label{eqone}
\end{equation}
where $N_K$, with $K=A,B,C,D$, is the number of events in region $K$
and $R^{\mathrm{bg}}=N_A^{\mathrm{bg}}\cdot
N_D^{\mathrm{bg}}/(N_B^{\mathrm{bg}}\cdot N_C^{\mathrm{bg}})$ is the
so-called background correlation, where $N_K^{\mathrm{bg}}$ with
$K=A,B,C,D$, is the a priori unknown number of background events in
each region.  The number of signal events in each of the three
background control regions ($N_K^{\mathrm{sig}}$) is taken into
account in Eq.~(\ref{eqone}) via the signal leakage fractions
$f_K\equiv N_K^{\mathrm{sig}}/N_A^{\mathrm{sig}}$ with $K=B,C,D$. MC
simulations of the signal are used to evaluate the signal leakage
fractions. Equation~(\ref{eqone}) with $R^{\mathrm{bg}}=1$ can be
understood as the relationship arising from the application of the
standard $ABCD$ method~\cite{pr:d83:052005} taking into account the
contributions from signal events in each region.
 
The ratio of the signal yield to the number of photon candidates in
region $A$ is used to estimate the signal purity and is above $90\%$
in all bins of the measured distributions. The signal yield is
extracted using leakage fractions from MC simulations and, hence,
depends on the modelling of the final state. This dependence is
studied by comparing the results obtained using either \pyt\ or \sher\
simulations. The calculations using the two generators lead to similar
signal yields and the differences are taken as systematic
uncertainties. The signal yields obtained using \pyt\ for the signal
leakage fractions are taken as the nominal ones. The distributions of
the signal yields as functions of $\etg$ in different regions of
$|\etag|$ are well described by the \pyt\ and \sher\ MC simulations,
except for $\etg\gtrsim 900$~\GeV\ in the range $|\etag|<1.37$. The
impact of this mismodelling in the measured cross section, which is
estimated by reweighting the MC simulations so that the distributions
have the same shape as the data and re-evaluating the unfolding
correction factors (see Section~\ref{fiducial}), is found to be
negligible.
 
For the nominal results it is assumed that the photon isolation and
identification variables are uncorrelated for background events, i.e.\
$R^{\mathrm{bg}}=1$. The dependence of the signal yield on this
assumption is investigated in validation regions, which are defined
within the background control regions $B$ and $D$. Region $B$ is
subdivided into two regions:  region $B^{\prime}$ of tight photon
candidates with $\etisoc +2$~\GeV~$<\etiso < \etisoc+10$~\GeV\ and
region $B^{\prime\prime}$ of tight photon candidates with $\etiso >
\etisoc+10$~\GeV. Likewise, region $D$ is subdivided into two regions,
$D^{\prime}$ and $D^{\prime\prime}$, using the same separation in
$\etiso$ as above. The four regions $B^{\prime}$, $B^{\prime\prime}$,
$D^{\prime}$ and $D^{\prime\prime}$ are used to extract values of
$R^{\mathrm{bg}}$ from the data after accounting for the signal
leakage fractions in those regions using either \pyt\ or \sher\ MC
simulations. The dependence on the signal leakage is investigated by
increasing the lower limits on $\etiso$ for the validation regions,
$\etisoc+2$~\GeV\ ($\etisoc+10$~\GeV), each time by $1$~\GeV\ up to
$\etisoc+7$~\GeV\ ($\etisoc+15$~\GeV) for regions $B^{\prime}$ and
$D^{\prime}$ ($B^{\prime\prime}$ and $D^{\prime\prime}$), keeping the
width in $\etiso$ fixed to $8$~\GeV\ for the regions $B^{\prime}$ and
$D^{\prime}$. The maximum deviations of $R^{\mathrm{bg}}$ from unity
in the validation regions vary in the range $15\%$--$40\%$ depending
on $\etg$ and $\etag$. The differences between the nominal signal
yields and those obtained using the maximum deviations of
$R^{\mathrm{bg}}$ from unity observed in the validation regions are
taken as systematic uncertainties.
 
Electrons or positrons can be misidentified as photons and represent
an additional source of background. This background is largely
suppressed by the photon selection. The residual background
contribution is evaluated using MC simulations of the Drell--Yan
processes $Z/\gamma^*\rightarrow e^+e^-$ and $W\rightarrow e\nu$ and
is found to be sub-percent in the phase-space region of the
analysis. Accordingly, no subtraction is performed and a systematic
uncertainty of the size of the evaluated background is assigned.
 
\section{Fiducial phase space and unfolding}
\label{fiducial}
The measured differential cross section is unfolded from the
distribution in $\etg$ separately for each of the four regions in
$|\etag|$ after the selection explained in Section~\ref{evesel} and
after background subtraction, as discussed in
Section~\ref{bacsub}. The phase-space region of the measurement
closely follows the applied event selection and is indicated in
Table~\ref{tablezero}. The fiducial phase-space region is defined at
particle level for all particles with a decay length of $c\tau>10$~mm;
these particles are referred to as `stable'. The isolation requirement
on the photon at particle level is based on the total transverse
energy of all stable particles, excluding muons and neutrinos, in a
cone of size $\Delta R=0.4$ around the photon direction after the
contributions from the  photon itself and the UE are subtracted. The
same subtraction procedure of the UE used at detector level is applied
at the particle level. The particle-level requirement on $\etiso$ is
also the same as the one used at detector level, i.e.\
$\etiso<\etisoc$.
 
The distributions of the background-subtracted signal yields as
functions of $\etg$ in different regions of $|\etag|$ are unfolded to
the particle level using the MC samples of events via a bin-by-bin
technique which corrects for resolution effects and the efficiency of
the photon selection through the formula
 
\begin{equation*}
\dgen(i)=\frac{N_A^{\mathrm{sig}}(i)\ C^{\mathrm{MC}}(i)}{{\cal L}\ \Delta \etg(i)},
\label{eqthree}
\end{equation*}
where $\dgenn(i)$ is the differential cross section in bin $i$,
$N_A^{\mathrm{sig}}(i)$ is the signal yield in bin $i$,
$C^{\mathrm{MC}}(i)$ is the unfolding correction factor in bin $i$,
${\cal L}$ is the integrated luminosity and $\Delta\etg(i)$ is the
width of bin $i$. The unfolding correction factors are computed using
the MC samples as
$C^{\mathrm{MC}}(i)=N^{\mathrm{MC}}_{\mathrm{part}}(i)/N^{\mathrm{MC}}_{\mathrm{det}}(i)$,
where $N^{\mathrm{MC}}_{\mathrm{part}}(i)$
($N^{\mathrm{MC}}_{\mathrm{det}}(i)$)  is the number of events
generated (reconstructed) in bin $i$ at the particle (detector)
level. The \pyt\ MC simulation is used in the evaluation of the
unfolding correction factors to obtain the nominal cross section. The
unfolding correction factors vary approximately between $1.1$ and
$1.3$ depending on $\etg$ and $|\etag|$. The dependence of the results
on the modelling of the final state, namely the inclusion of
higher-order tree-level matrix elements, the parton shower approach
and the hadronisation model, is studied with the \sher\ MC
simulation. The differences between the results obtained by using
either the \pyt\ or  \sher\ MC simulations are taken as systematic
uncertainties. The cross section is also obtained using an iterative
Bayesian unfolding method~\cite{nim:a362:487} and compared with the
nominal result; the two results are consistent with each other
independently of whether the \pyt\ or \sher\ MC simulations are used
for the unfolding.
 
\section{Systematic uncertainties}
\label{syst}
The sources of systematic uncertainty that affect the measurement are
discussed below. These sources include the signal modelling, the
background subtraction, the photon identification and isolation, the
unfolding procedure, the modelling of pile-up, the trigger efficiency,
the luminosity measurement and the photon energy scale and
resolution. For some of the systematic uncertainties, the Bootstrap
technique~\cite{ISBN:978-3-935702-41-6} is used to evaluate the data
statistical influence on the uncertainties. The results are then used
in a fit to smooth the systematic uncertainties.
 
{\textbf{Model dependence of the signal leakage fractions.}}
The nominal signal leakage fractions used in the extraction of the
signal yield are evaluated with the \pyt\ simulation.  The effect of
the implementation of the matrix-element calculation in the generator
as well as the models used for the parton shower and hadronisation are
evaluated by comparing the nominal results with those obtained using
the \sher\ simulation for the determination of the signal leakage
fractions (see Section~\ref{bacsub}). The differences are taken as
systematic uncertainties. The resulting uncertainty in the measured
cross section is always less than $2\%$ and typically less than $1\%$.
 
{\textbf{Background subtraction.}}
The data-driven background subtraction depends on the choice of
background control regions. This dependence is studied by varying the
lower limit on $\etiso$ for regions $B$ and $D$ by $\pm 1$~\GeV,
removing the upper limit on $\etiso$ for regions $B$ and $D$, as well
as changing the choice of inverted photon identification
variables. For the last variation, the analysis is repeated increasing
or decreasing the number of the shower-shape variables computed from
the energy deposits in the first layer of the EM calorimeter that are
removed from the list of tight requirements used in the classification
of non-tight photon candidates. For each variation, the differences
between the nominal signal yields and those extracted using the
modified background control regions are taken as systematic
uncertainties. The resulting uncertainties in the measured cross
section due to the first two variations mentioned above are less than
$0.2\%$. The third variation leads to uncertainties in the measured
cross section that are less than $2\%$ and typically less than
$1\%$. As described in Section~\ref{bacsub}, a systematic uncertainty
is assigned to the assumption $R^{\mathrm{bg}}=1$. The resulting
uncertainty in the measured cross section is less than $2.5\%$ and
typically less than $1\%$. The background contribution from electrons
or positrons misidentified as photons is sub-percent and a systematic
uncertainty is included by assigning the full size of this background;
as an example, the background from $W$ ($Z$) decays is $0.22\%$
($0.07\%$) for $|\etag|<0.6$.
 
{\textbf{Photon identification and isolation.}}
The uncertainties in the efficiency scale factors, which are applied
to simulated events to match the tight photon identification
efficiency measured in data~\cite{epj:c79:205}, are propagated to the
measured cross section. The resulting uncertainty in the measured
cross section is in the range $1\%$--$3\%$. The uncertainties in the
cross section due to the modelling of $\etiso$ in simulated events are
evaluated by comparing the nominal results with those obtained using
MC samples of events in which the data-driven correction to $\etiso$
is not applied (see Section~\ref{evesel}). The resulting uncertainty
in the measured cross section is less than $2\%$ except for $\etg>
500$~\GeV\ in the regions $0.6<|\etag|<1.37$ and $1.56<|\etag|<1.81$,
where it increases to $3\%$.
 
{\textbf{Model dependence of the unfolding.}}
The nominal unfolding correction factors are computed using the \pyt\
simulation. As discussed in Section~\ref{fiducial}, the effects on the
unfolding correction factors due to the matrix elements and the parton
shower and hadronisation models employed in the generators are
investigated. The difference in the cross section between the nominal
result and that obtained using the \sher\ simulation for the
determination of the unfolding correction factors is taken as a
systematic uncertainty. The uncertainties in the measured cross
section are less than $2\%$ and typically less than $1\%$. The
uncertainty in the cross section due to the statistical uncertainty of
the MC samples is propagated into the measured cross section and
typically amounts to less than $0.5\%$.
 
{\textbf{Pile-up.}}
A variation in the pile-up reweighting of simulated events is included
to cover the uncertainty in the ratio of the predicted to measured
inelastic cross sections~\cite{prl:117:182002}. The resulting
uncertainty in the measured cross section is less than $1.5\%$.
 
{\textbf{Trigger efficiency.}}
The uncertainty in the trigger efficiency is estimated using the same
methodology as in Ref.~\cite{epj:c77:317} and is propagated into the
measured cross section. The uncertainty in the measured cross section
is at most $0.4\%$.
 
{\textbf{Luminosity measurement.}}
The uncertainty in the integrated luminosity is $2.1\%$. It is fully
correlated between all bins with $\etg>150$~\GeV\ of the cross
section. For the bin with $125<\etg<150$~\GeV\ only data from the 2015
running period is used and the uncertainty in the luminosity
measurement is not fully correlated with the uncertainty in the
combined integrated luminosity corresponding to the 2015 and 2016
running periods. However, the impact of the uncorrelated part can be
safely neglected.
 
{\textbf{Photon energy scale and resolution.}}
A detailed analysis of the sources of uncertainty in the energy scale
and resolution for photons was made with Run 1
data~\cite{epj:c74:3071}. The same model is implemented and updated
for Run 2 data~\cite{JINST14:P03017}. The sources of uncertainty
include: the uncertainty in the overall energy scale adjustment using
$Z\rightarrow e^+e^-$ decays; the uncertainty in the non-linearity of
the energy measurement at the cell level; the uncertainty in the
relative calibration of the different calorimeter layers; the
uncertainty in the amount of material in front of the calorimeter; the
uncertainty in the modelling of the reconstruction of photon
conversions; the uncertainty in the modelling of the lateral shower
shape; the uncertainty in the modelling of the sampling
term;\footnote{The relative energy resolution is parameterised as
$\sigma(E)/E = a/\sqrt{E} \oplus b/E \oplus c$, where $a$ is the
sampling term, $b$ is the noise term and $c$ is the constant term.}
the uncertainty in the measurement of the constant term in $Z$-boson
decays. The sources of uncertainty are modelled using independent
components to account for their $\eta$ dependence. A total of $76$
individual components influencing the energy scale and resolution of
the photon are identified to assess the overall uncertainty in the
energy measurement. All the uncertainty components are propagated
separately through the analysis to keep track of the information about
the correlations between different bins. The systematic uncertainty in
the measured cross section is evaluated by varying each individual
source of uncertainty separately by $\pm 1\sigma$ in the MC
simulations and then adding the uncertainty contributions in
quadrature. The resulting uncertainties in the measured cross section
are typically less than $0.5\%$ for the energy resolution and in the
range $1\%$--$16\%$ for the energy scale.
 
\begin{figure}[p]
\setlength{\unitlength}{1.0cm}
\begin{picture} (18.0,16.0)
\put (0.0,8.0){\includegraphics[width=8cm]{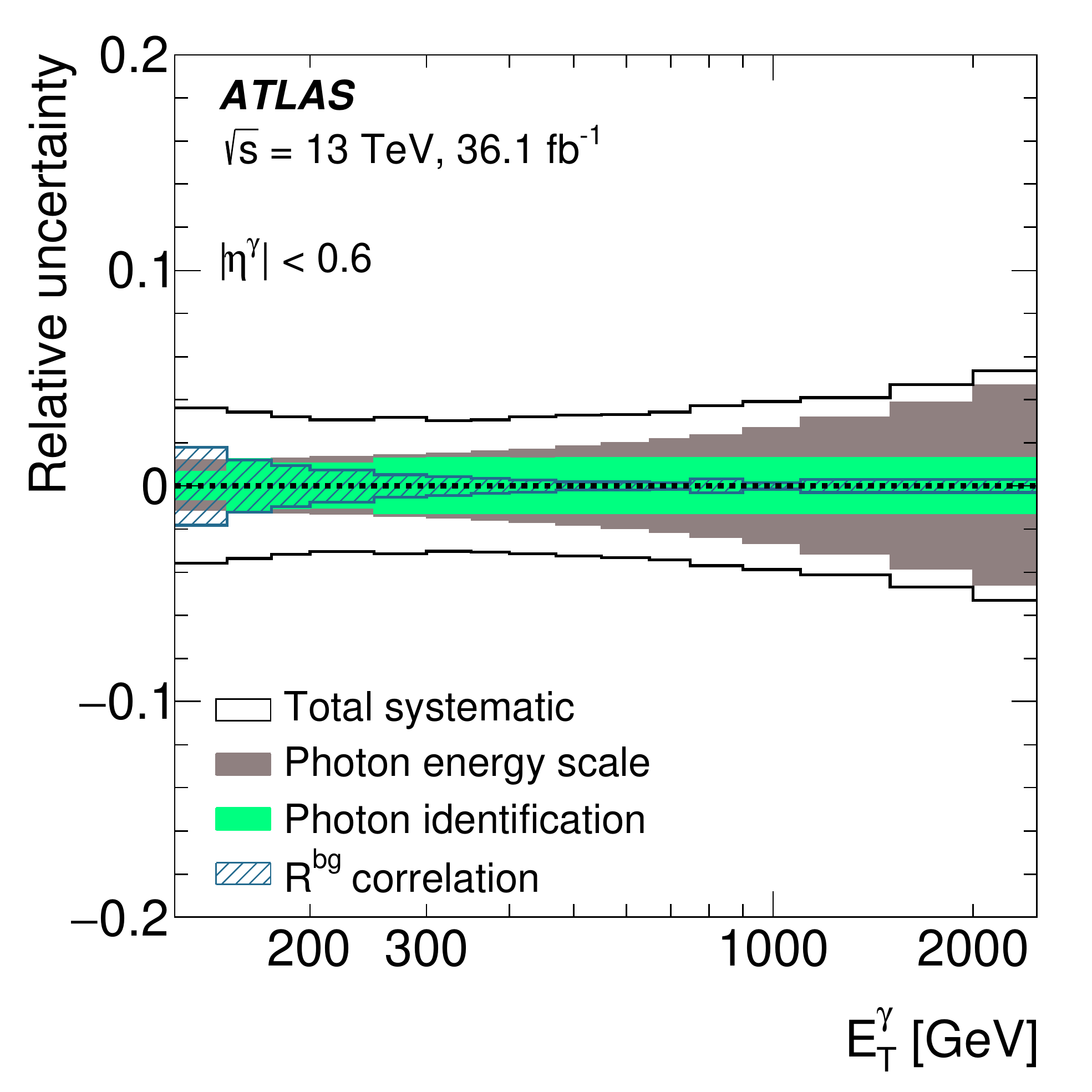}}
\put (8.0,8.0){\includegraphics[width=8cm]{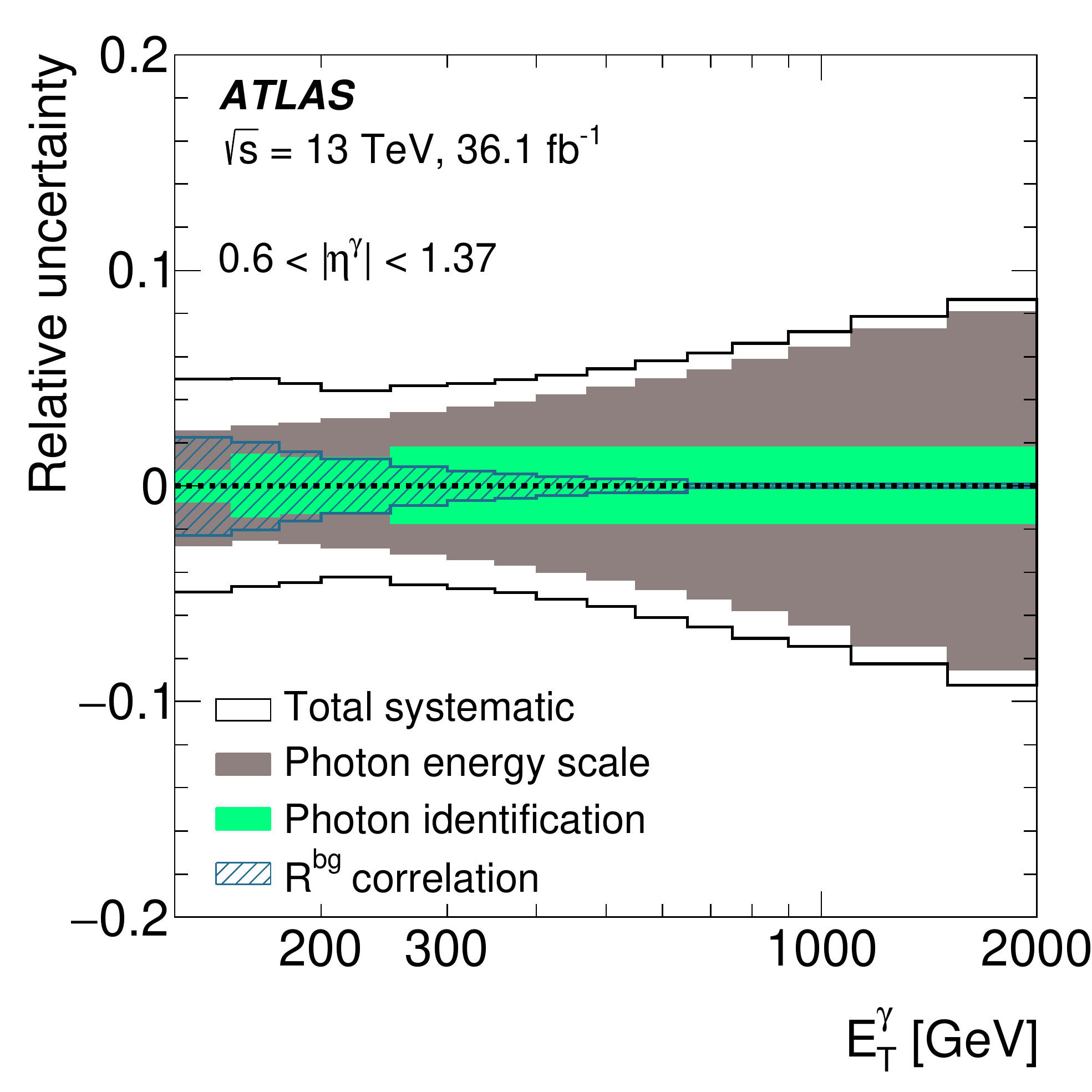}}
\put (0.0,0.0){\includegraphics[width=8cm]{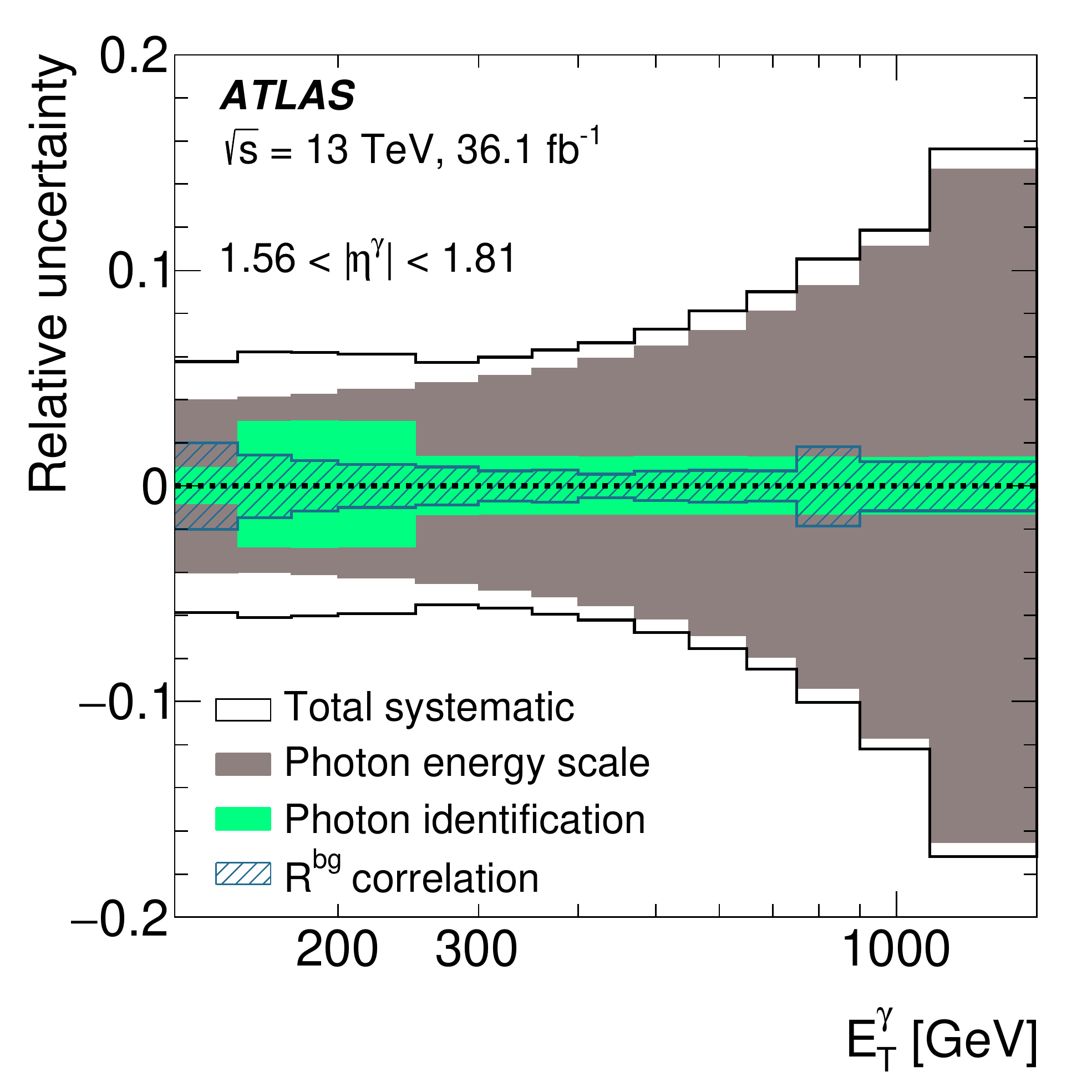}}
\put (8.0,0.0){\includegraphics[width=8cm]{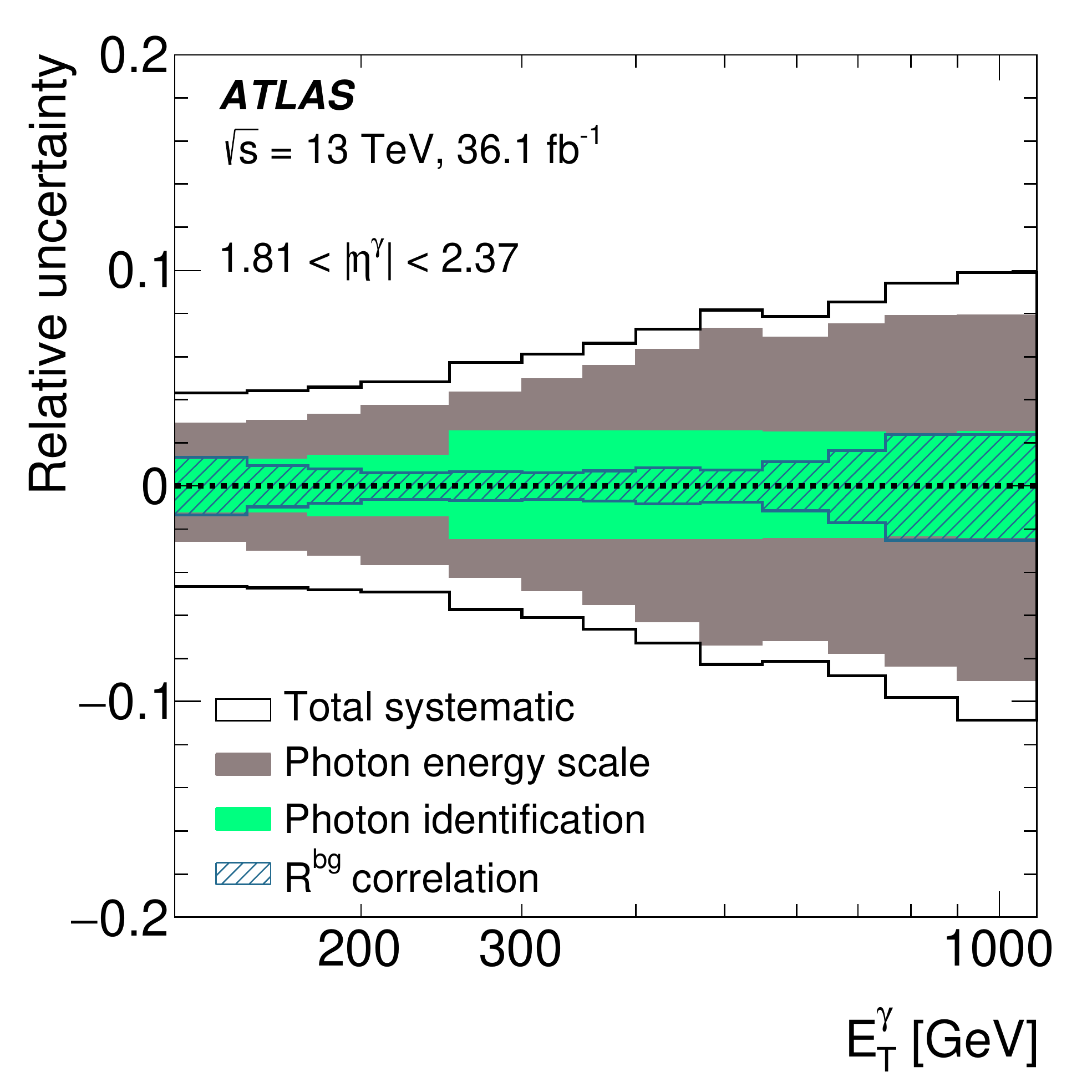}}
\end{picture}
\caption
{
The relative total systematic uncertainty in the cross section
(white areas) as a function of $\etg$ in different $|\etag|$
regions. The relative uncertainty due to the photon energy scale
(grey areas), the relative uncertainty due to the photon
identification efficiency (green areas) and the relative uncertainty
due to $R^{\mathrm{bg}}$ (blue hatched areas) are also shown.
}
\label{fig1}
\end{figure}

\begin{figure}[p]
\setlength{\unitlength}{1.0cm}
\begin{picture} (18.0,16.0)
\put (0.0,8.0){\includegraphics[width=8cm]{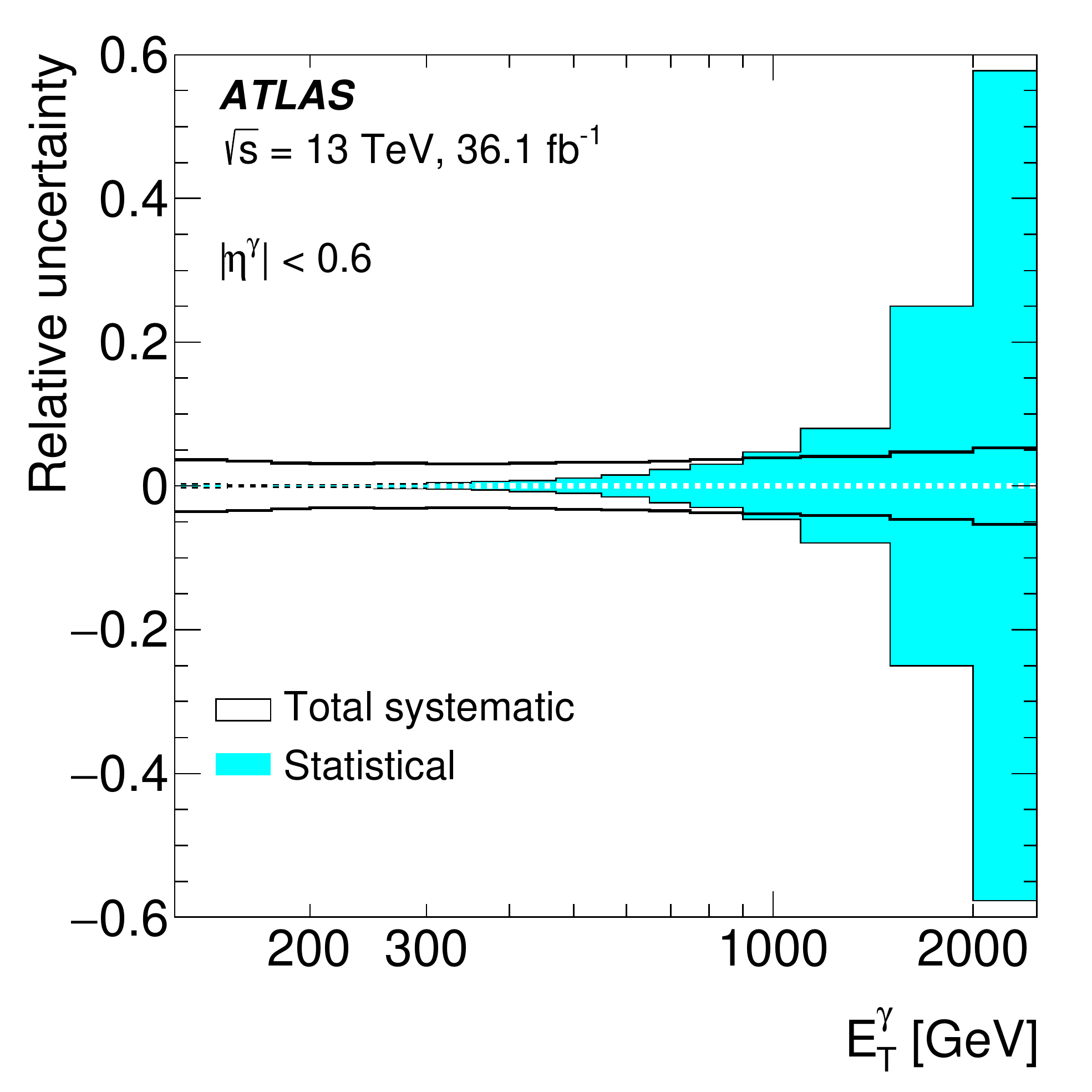}}
\put (8.0,8.0){\includegraphics[width=8cm]{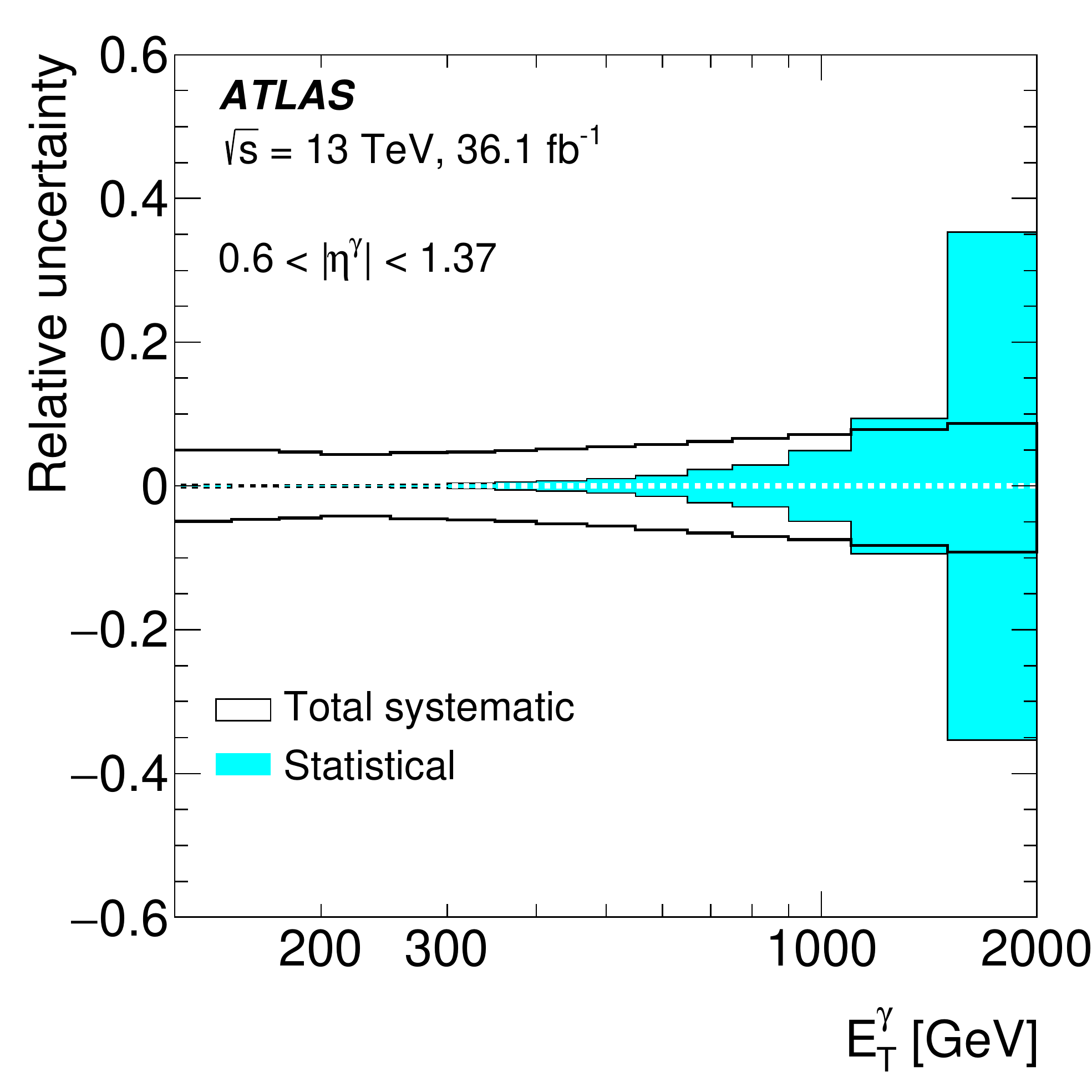}}
\put (0.0,0.0){\includegraphics[width=8cm]{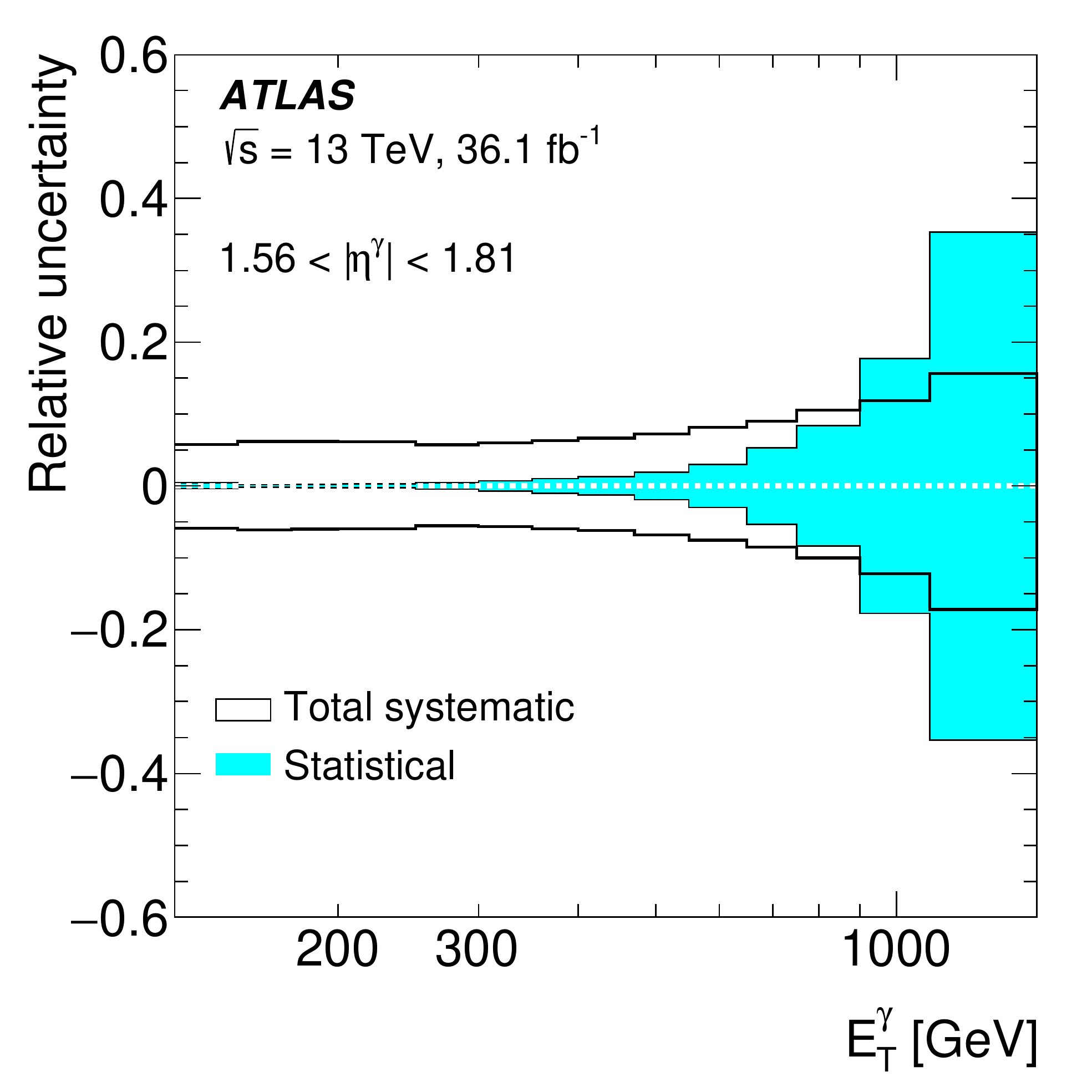}}
\put (8.0,0.0){\includegraphics[width=8cm]{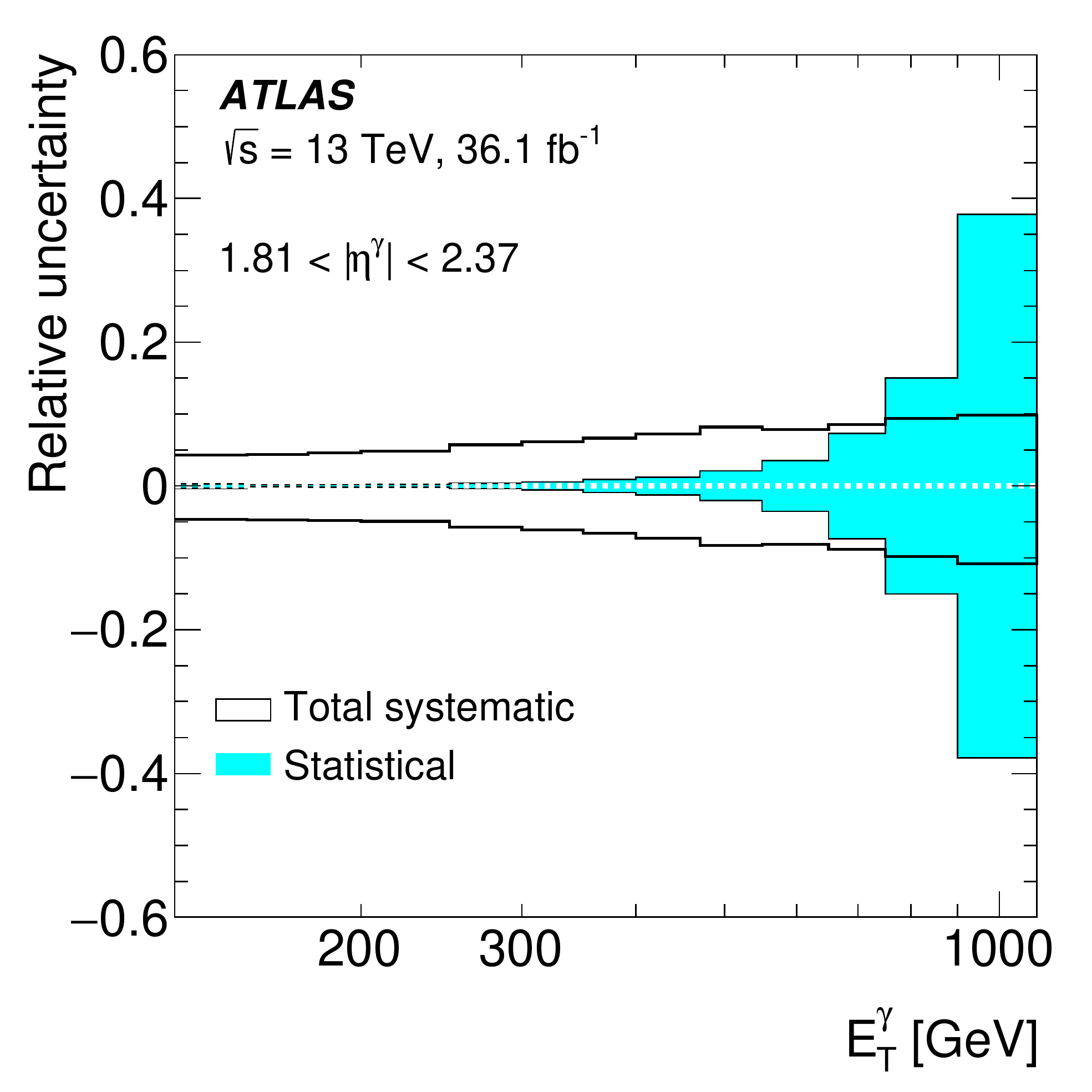}}
\end{picture}
\caption
{
The relative total systematic uncertainty in the cross section
(white areas) as a function of $\etg$ in different $|\etag|$
regions. The relative statistical uncertainty from the data is also
shown (cyan areas).
}
\label{fig2}
\end{figure}

{\textbf{Total systematic uncertainty.}}
The total experimental systematic uncertainty is calculated by adding
in quadrature the uncertainties listed above. The total systematic
uncertainty is in the range $3\%$--$17\%$, depending on $\etg$ and the
$|\etag|$ region. The dominant sources of uncertainty arise from the
photon energy scale, the photon identification efficiency and the
integrated luminosity. Figure~\ref{fig1} shows the total systematic
uncertainty for each measured cross section together with the dominant
components. The systematic uncertainty dominates the total
experimental uncertainty for $\etg\lesssim 1$~\TeV\ and
$|\etag|<1.37$, whereas for higher $\etg$ values, the statistical
uncertainty of the data limits the precision of the measurement, as
can be seen in Figure~\ref{fig2}. This represents an improvement
relative to the previous measurement~\cite{pl:b770:473}, where the
region dominated by the systematic uncertainties was $\etg\lesssim
600$~GeV.
 
\section{Theoretical predictions}
\label{nlo}
The NLO pQCD predictions are computed using two programs, namely
\jetp~1.3.1\_2 and \sher~2.2.2. The \jetp\ program provides NLO QCD
calculations of the direct and fragmentation contributions to the
prompt-photon cross section. The number of massless quark flavours is
set to five. The renormalisation ($\muR$), factorisation ($\muF$) and
fragmentation ($\muf$) scales are chosen to be
$\muR=\muF=\muf=\etg$. Variations of these scales are considered to
get an estimate of the uncertainties due to missing higher-order terms
in the perturbative expansion (see Section~\ref{nlounc}). The nominal
prediction is obtained using the MMHT2014~\cite{epj:c75:204} PDF set
and the BFG set II of parton-to-photon fragmentation
functions~\cite{epj:c2:529}, both of which are determined at NLO. The
strong coupling constant is set to $\asz=0.120$ since it is the value
assumed in the MMHT2014 PDF fit. For the electromagnetic coupling
($\alpha_{\textsc{em}}$), the low-energy limit of $1/137.036$ is
used. A parton-level isolation criterion is used which requires the
total transverse energy from the partons inside a cone of size $\Delta
R=0.4$ around the photon direction to be below the same $\etisoc$
applied at particle and detector level; this isolation prescription is
referred to as `fixed-cone isolation'. Predictions based on other PDF
sets, namely CT14~\cite{pr:d93:033006}, ABMP16~\cite{pr:d96:014011},
HERAPDF2.0~\cite{epj:c75:580} and NNPDF3.0~\cite{jhep:1504:040}, are
also compared with the data; in each case $\asz$  is set to the value
assumed in the PDF fit.
 
The \sher~2.2.2 program consistently combines parton-level
calculations of $\gamma+ {\textrm{(1,\ 2)-jet}}$ events at NLO in pQCD
and $\gamma+{\textrm{(3,\ 4)-jet}}$ events at
LO~\cite{jhep:0202:044,prl:108:111601} supplemented with a parton
shower~\cite{jhep:0803:038} while avoiding double-counting
effects~\cite{jhep:1304:027}. The sample of events generated with
\sher~2.2.2 is different from the sample of events from \sher~2.1.1
described in Section~\ref{datmc}. The former includes the higher-order
virtual corrections for $\gamma+ {\textrm{(1,\ 2)-jet}}$ and is not
passed through the ATLAS detector simulation programs since the goal
is to compare the \sher~2.2.2 prediction with the measured cross
section. Photon isolation at the matrix-element level is applied by
using Frixione's criterion with ${\cal R}=0.1$, $n=2$ and
$\epsilon=0.1$ (see Section~\ref{datmc}). The parameters are chosen to
be as loose as possible to minimise a bias from the application of the
photon isolation requirement at particle level. Since Frixione's
criterion requires the upper limit on the transverse energy isolation
to be exactly zero at ${\mathfrak r}=0$, the criterion cannot be
strictly looser than any non-zero photon isolation requirement at
detector or particle level for all ${\mathfrak r}<{\cal R}$. The
photon isolation requirement is applied (see Table~\ref{tablezero})
using the procedure described in Section~\ref{fiducial}; the
prescription employed is referred to as  `hybrid-cone
isolation'~\cite{jphys:g44:044007,1904.01044} since it includes the
application of the Frixione's criterion at a small value of $\Delta R$
(${\cal R}=0.1$) and the fixed-cone isolation at $\Delta R=0.4$ used
for the fiducial region of the measurement. Dynamic factorisation and
renormalisation scales are adopted ($\etg$) as well as a dynamical
merging scale  with $\bar
Q_{\mathrm{cut}}=20$~\GeV~\cite{jphys:g44:044007}.  The strong
coupling constant is set to $\asz=0.118$.  The same prescription for
the electromagnetic coupling as for the \jetp\ prediction is
used. Fragmentation into hadrons and simulation of the UE are
performed using the same models as for the \sher\  samples of
simulated events presented in Section~\ref{datmc}. The NNLO NNPDF3.0
PDF set is used in conjunction with the corresponding \sher\ tuning.
 
The NNLO QCD prediction is calculated~\cite{1904.01044} in the
\nnlojet\ framework. The NNLO corrections include three types of
parton-level contributions, namely the two-loop corrections to
photon-plus-one-parton production, the virtual corrections to
photon-plus-two-parton production and the tree-level
photon-plus-three-parton production. Only direct-photon processes are
included; fragmentation processes are circumvented by the application
of the Frixione's criterion at small $\Delta R$ using the same
parameter settings as for the \sher~2.2.2 prediction. The photon
transverse energy is used as the baseline choice for the
renormalisation and factorisation scales. The NNPDF3.1 PDF
set~\cite{epj:c77:663} at NNLO is used. The electromagnetic coupling
is taken in the $G_{\mu}$-scheme~\cite{jhep:01:060} as
$\alpha_{\textsc{em}}^{G_{\mu}}=1/132.232$. Photon isolation is
implemented following the `hybrid-cone isolation' prescription
described above. For the \nnlojet\ prediction the fixed-cone isolation
at $\Delta R=0.4$ is applied at the parton level using the same
requirement as for the particle level. The prediction at NLO pQCD in
the \nnlojet\ framework is also calculated to illustrate the
improvements achieved by including the NNLO QCD corrections. The NLO
QCD prediction of \nnlojet\ differs from that of \jetp\ in the
fragmentation contribution, the photon isolation prescription, the
value of $\alpha_{\textsc{em}}$ and the PDF set.
 
There are several differences between the calculations using \jetp,
\sher~2.2.2 and \nnlojet: the calculations from \nnlojet\ include NNLO
QCD corrections; the calculations using \sher~2.2.2 include
higher-order contributions as well as parton showers; the application
of the Frixione's criterion (\sher~2.2.2 and \nnlojet) at
matrix-element level allows the fragmentation contribution to be
ignored; and the prediction for the cross section using \sher~2.2.2 is
at particle level and include UE effects. A compilation of the major
features of the three different approaches is shown in
Table~\ref{tableone}.
 
\begin{table}[h]
\caption{
Major features of the three predictions used for inclusive
isolated-photon production.
}
\setlength{\unitlength}{1.0cm}
\begin{picture} (18.0,3.0)
\put (0.0,1.0){\scalebox{1.0}{\centerline{
\begin{tabular}{|l|l|l|l|l|l|l|} \hline
Program      & Order in $\as$ &  Fragmentation & Parton  & Isolation & $\alpha_{\textsc{em}}$ & Particle \\
&                        &                          & shower &                &                                        & Level \\
\hline
\jetp\          & NLO                &  Yes &  No &  Fixed cone & $1/137.036$ & No \\ \hline
\sher~2.2.2 & NLO for  $\gamma+ {\textrm{(1,\ 2)-jet}}$             &  No &  Yes &  Hybrid cone & $1/137.036$ & Yes \\
& LO for $\gamma+{\textrm{(3,\ 4)-jet}}$                 &       &        &       &    & \\ \hline
\nnlojet\     & NNLO             &  No &  No &  Hybrid cone & $1/132.232$ & No \\ \hline
\end{tabular}
}}}
\end{picture}
\label{tableone}
\end{table}
 
Electroweak corrections are not included in the calculations of \jetp,
\sher~2.2.2 or \nnlojet. As mentioned earlier, the corrections are
likely to play an important role at the \TeV\ scale. Electroweak
Sudakov corrections for single photon production are
available~\cite{pr:d92:073011}. However, these constitute only part of
the full electroweak corrections since the real corrections are not
included. The full electroweak corrections can differ significantly
from the  Sudakov corrections because the latter corrections and the
real corrections  partially cancel out when the presence of $W$ or $Z$
bosons in the final state is not vetoed. The measurement presented
here does not preclude the presence of massive gauge bosons in
addition to the high-transverse-momentum photon.
 
\subsection{Theoretical uncertainties in the NLO QCD predictions}
\label{nlounc}
The major source of uncertainty affecting the predictions is due to
terms beyond NLO in QCD. In the case of \jetp\ it is evaluated by
repeating the calculations using values of $\muR$, $\muF$ and $\muf$
scaled by the factors $0.5$ and $2$. The  three scales are either
varied simultaneously or independently; in addition, configurations in
which one scale is fixed and the other two are varied simultaneously
are also considered.  In all cases, the condition
$0.5\leq\mu_A/\mu_B\leq 2$ is imposed, where  $A,B={\mathrm{R,\ F,\
f}}$. The final uncertainty is taken as the largest deviation from
the nominal value among the $14$ possible variations; this is done
separately for the positive and negative contributions, so the final
uncertainty is not necessarily symmetric. In the case of \sher~2.2.2,
which does not include the fragmentation contribution, $\muR$ and
$\muF$ are varied as above and the largest deviation from the nominal
prediction among the six possible variations is taken as the
uncertainty. The resulting uncertainties in the prediction are
$10\%$--$15\%$ ($20\%$--$30\%$) for \jetp\ (\sher~2.2.2). The fact
that the uncertainties are larger for \sher\ than for \jetp\ is
attributed to multi-jet configurations which are accounted for only in
the case of \sher.
 
The uncertainty in the prediction of \jetp\ due to the uncertainty in
the proton PDFs is estimated by repeating the calculations using the
$50$ sets from the MMHT2014 error analysis and applying the Hessian
method~\cite{pr:d65:014013}. In the case of \sher~2.2.2, it is
estimated using the $100$ replicas from the NNPDF3.0 analysis. The
resulting uncertainties in the prediction is in the range $1\%$--$6\%$
for \jetp\ and $1\%$--$9\%$ for \sher~2.2.2, depending on $\etg$ and
the $|\etag|$ region.
 
The uncertainty in the prediction of \jetp\ (\sher~2.2.2) due to the
uncertainty in $\as$ is evaluated by repeating the calculations using
two additional sets of proton PDFs from the MMHT2014 (NNPDF3.0)
analysis for which different values of $\as$ at $\mz$ are assumed in
the fits, namely $0.118$ ($0.117$) and $0.122$ ($0.119$); in this way,
the correlation between $\as$ and the PDFs is preserved. The
uncertainties in \jetp\ associated with $\as$ are evaluated as the
differences between the varied and the nominal cross sections scaled
by a factor $1.5/2$. This scaling factor accounts for the difference
in $\asz$ between the nominal and the $\asz$-varied calculations
($0.002$) and the uncertainty in $\asz$ ($0.0015$) recommended by the
authors of MMHT2014~\cite{epj:c75:435}. The uncertainties in
\sher~2.2.2 due to $\as$ are evaluated as the differences between the
varied and the nominal cross sections scaled by a factor $1.5$ to
match the prescription used for MMHT2014. The resulting uncertainties
in the prediction of \jetp\ (\sher~2.2.2) are less than $3\%$ ($6\%$
except for the highest measured $\etg$ point in the region
$1.81<|\etag|<2.37$, where it increases to $12\%$).
 
Since the NLO pQCD prediction from \jetp\ is at the parton level while
the measurement is at the particle level, correction factors for the
non-perturbative (NP) effects of hadronisation and the underlying
event are needed for the former to match the data. However, the data
are corrected for pile-up and UE effects and the distributions are
unfolded to a phase-space definition in which the requirement on
$\etiso$ at particle level is applied after subtraction of the
UE. Therefore, corrections for NP effects are expected to be close to
unity. They are evaluated by computing the ratio of the particle-level
cross section for a \pyt\ sample with UE effects to the parton-level
cross section without UE effects. The resulting corrections are
consistent with unity within $\pm 1\%$. Thus, no correction is applied
to the NLO QCD prediction of \jetp\ and an uncertainty of $1\%$ is
assigned. The \sher~2.2.2 prediction is based on particle-level
observables after applying the requirements listed in
Table~\ref{tablezero} and, therefore, can be compared directly with
the measurement.  Although no tunes are available to assess the impact
of hadronisation and UE in the \sher~2.2.2 prediction, it is expected
that the associated uncertainty is of a similar size as that evaluated
using \pyt\ and, therefore, can be neglected in comparison to the
other uncertainties.
 
\begin{figure}[h]
\setlength{\unitlength}{1.0cm}
\begin{picture} (18.0,16.0)
\put (0.0,8.0){\includegraphics[width=8cm]{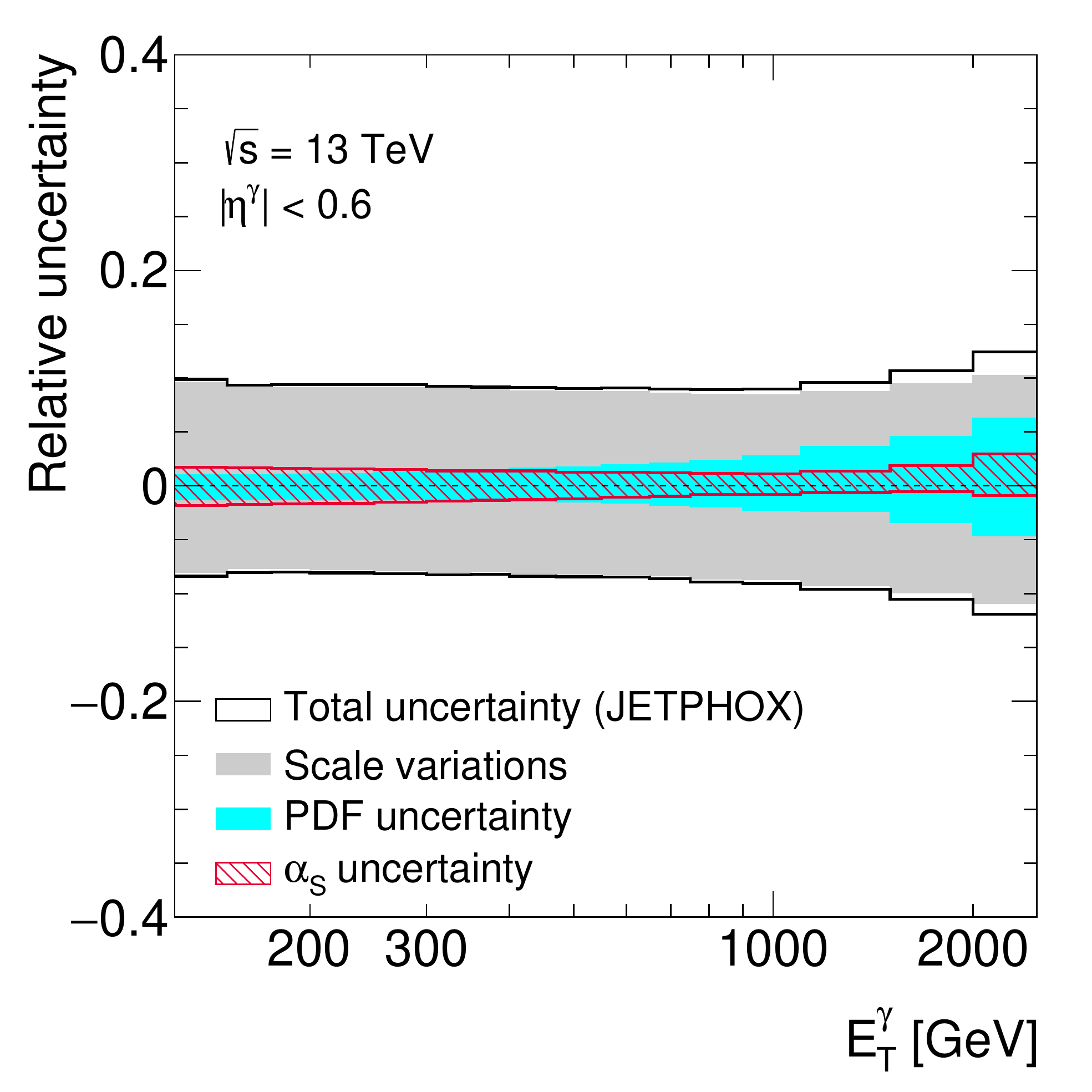}}
\put (8.0,8.0){\includegraphics[width=8cm]{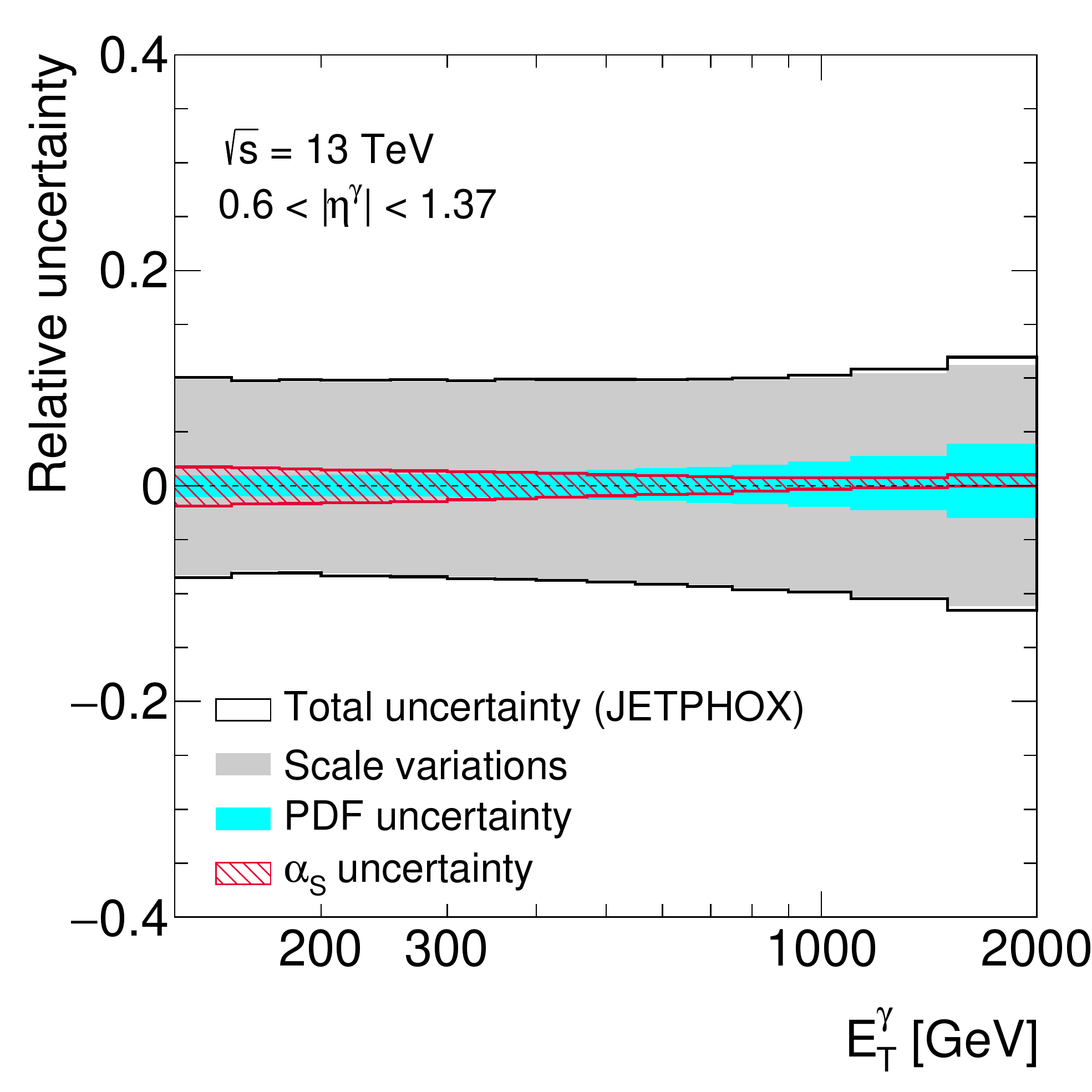}}
\put (0.0,0.0){\includegraphics[width=8cm]{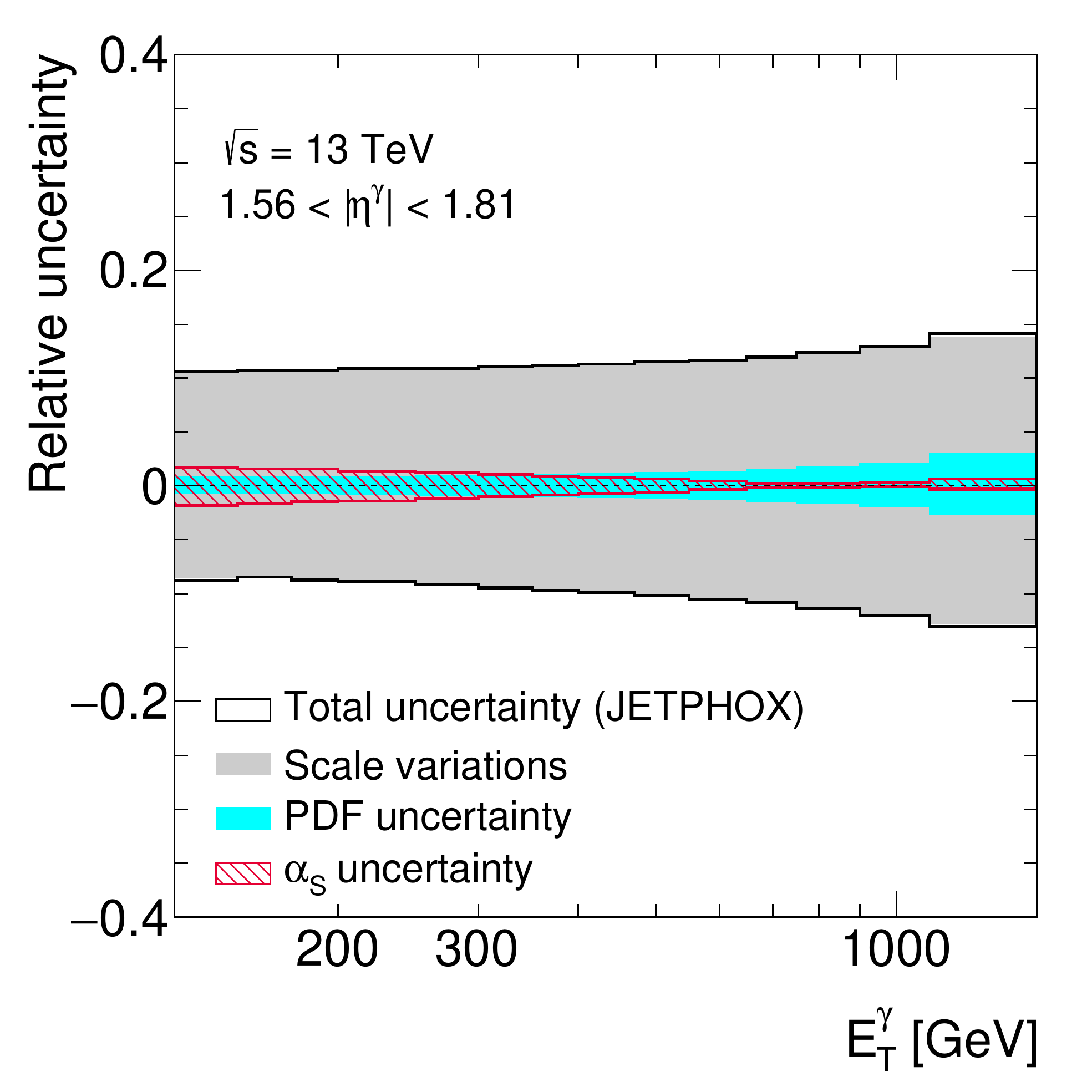}}
\put (8.0,0.0){\includegraphics[width=8cm]{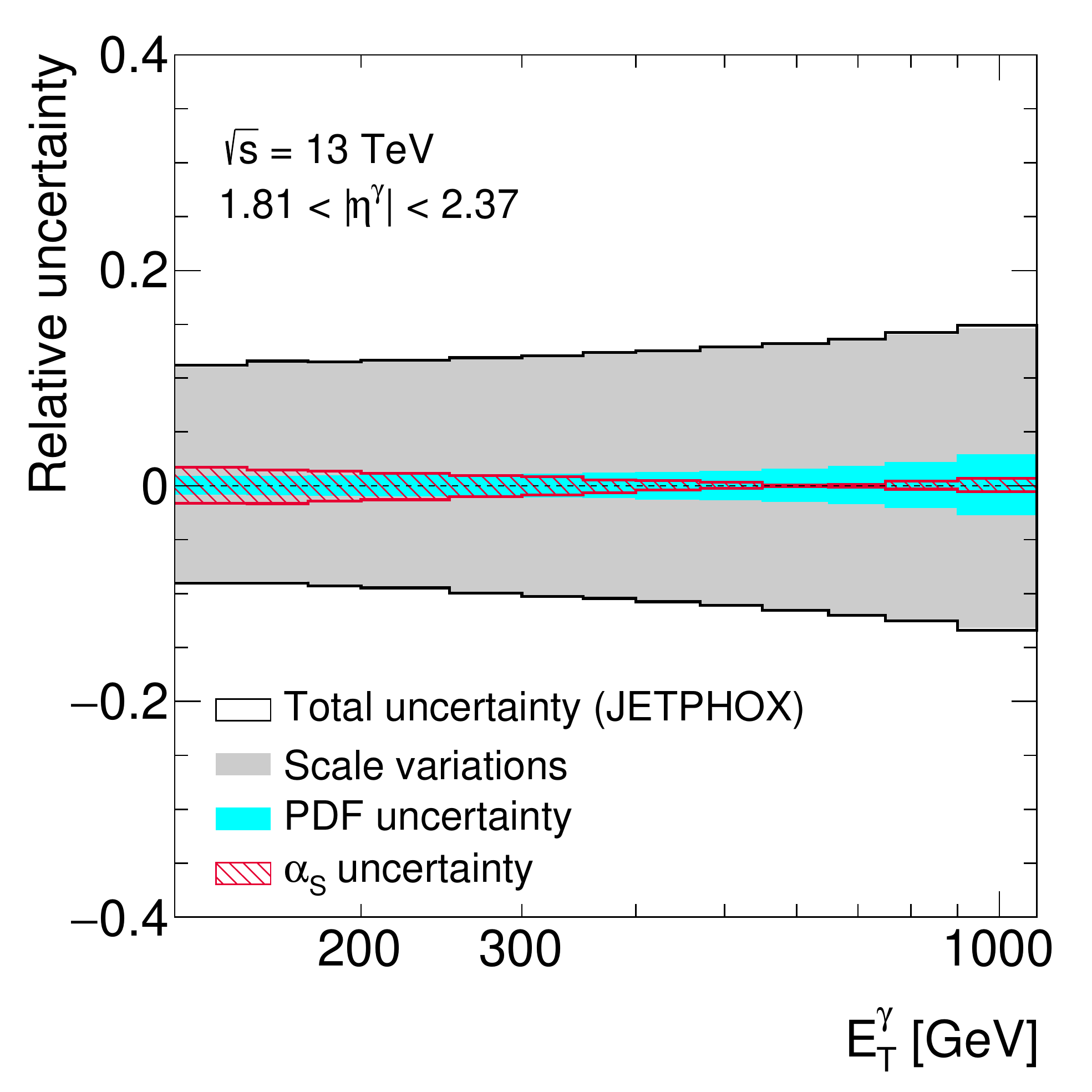}}
\end{picture}
\caption
{
The relative total theoretical uncertainty in the cross-section
prediction of \jetp\ as a function of $\etg$ in different regions of
$|\etag|$. The uncertainty from the scale variations, the
uncertainty from the PDFs and the uncertainty from the value of
$\as$ are also included.
}
\label{fig3}
\end{figure}

\begin{figure}[p]
\setlength{\unitlength}{1.0cm}
\begin{picture} (18.0,16.0)
\put (0.0,8.0){\includegraphics[width=8cm]{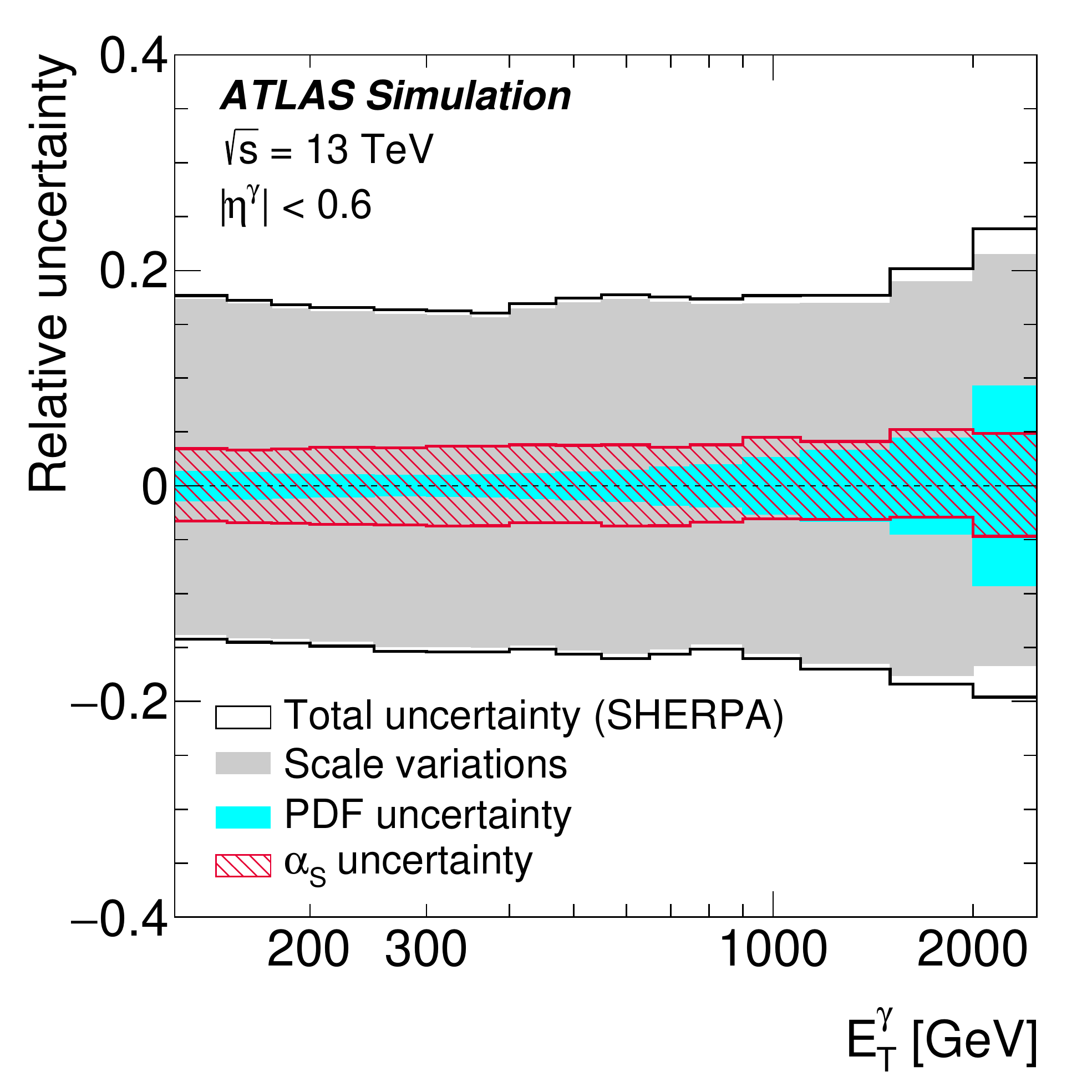}}
\put (8.0,8.0){\includegraphics[width=8cm]{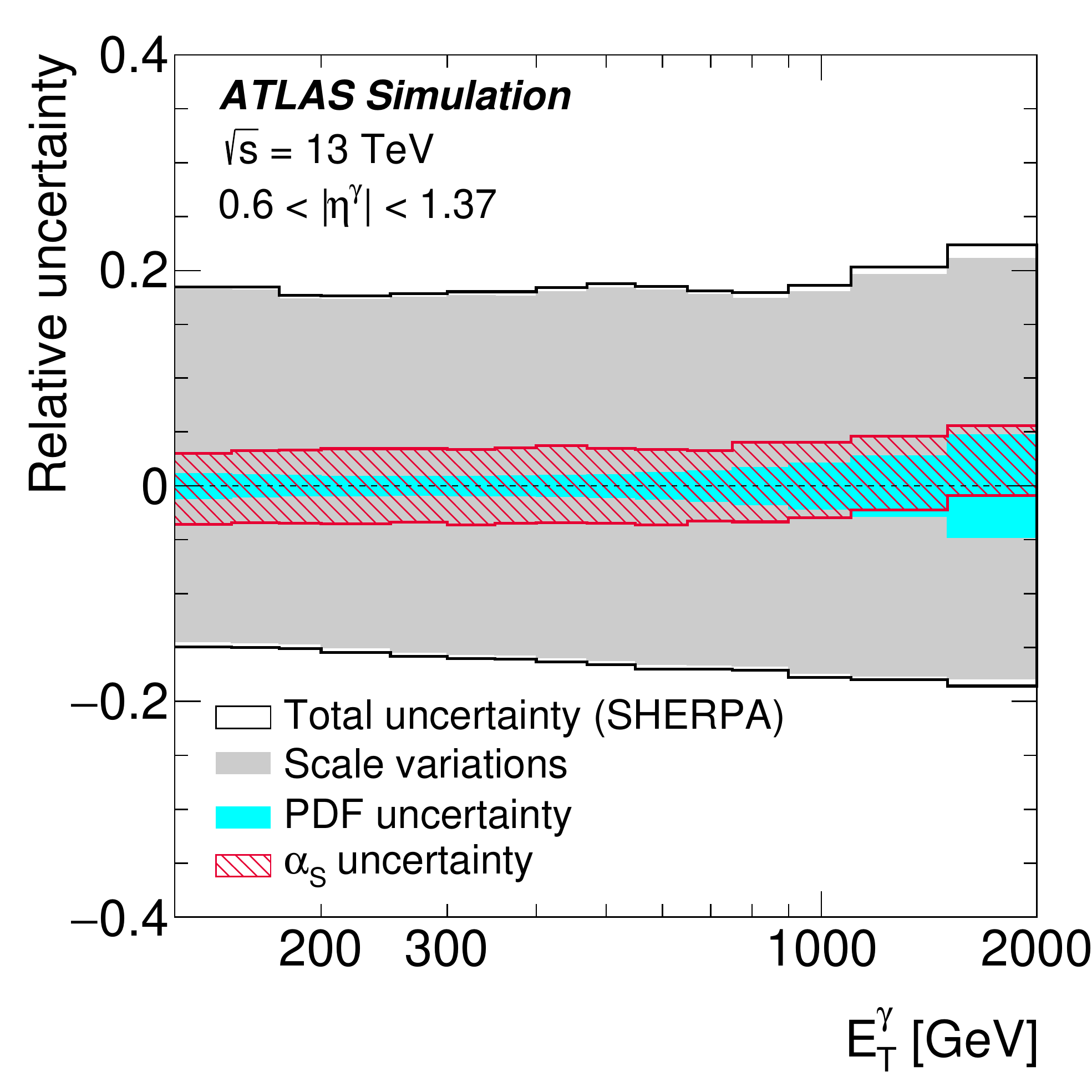}}
\put (0.0,0.0){\includegraphics[width=8cm]{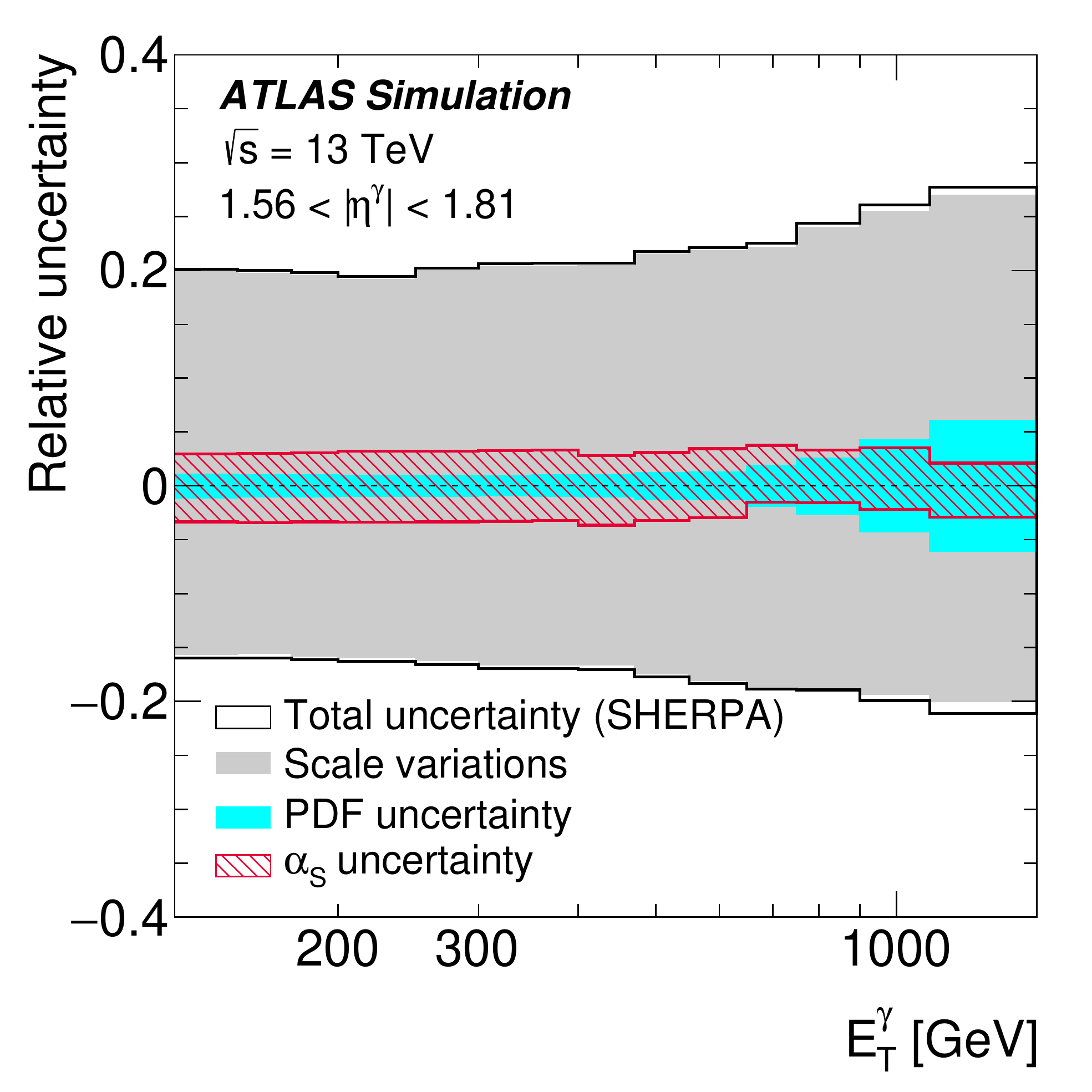}}
\put (8.0,0.0){\includegraphics[width=8cm]{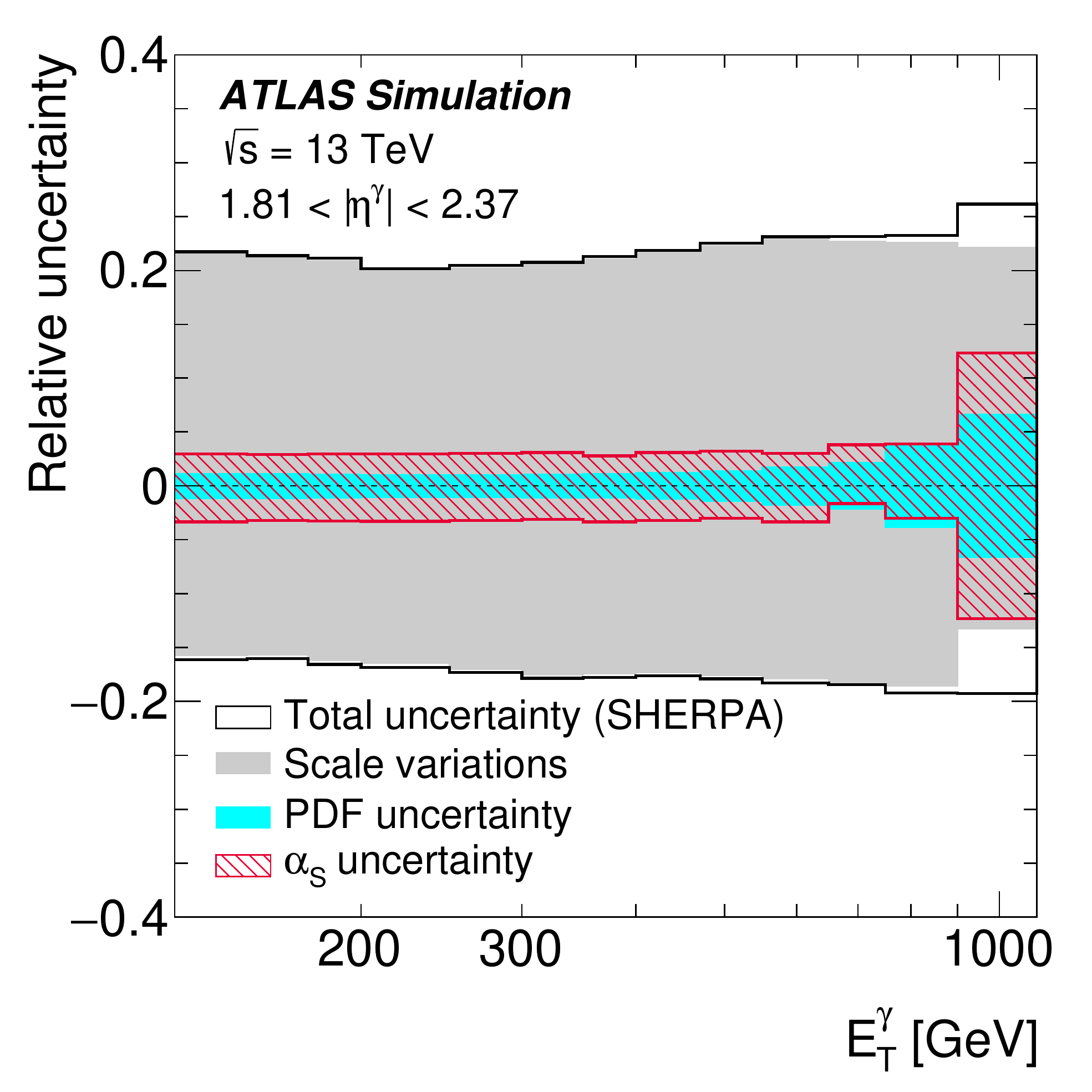}}
\end{picture}
\caption
{
The relative total theoretical uncertainty in the cross-section
prediction of \sher~2.2.2 as a function of $\etg$ in different
regions of $|\etag|$. The uncertainty from the scale variations, the
uncertainty from the PDFs and the uncertainty from the value of
$\as$ are also included.
}
\label{fig4}
\end{figure}

The total theoretical uncertainty is obtained by adding in quadrature
the individual uncertainties listed above (see Figures~\ref{fig3} and
\ref{fig4}). The dominant theoretical uncertainty for the prediction
using \jetp\ or \sher~2.2.2 is that arising from the variation of the
scales. The uncertainty arising from the PDFs ($\asz$) is the second
dominant uncertainty for the prediction of \jetp\ (\sher~2.2.2). A
prediction from \jetp\ using as scales $\muR=\muF=\muf=\etg/2$ is also
made and the associated theoretical uncertainties are calculated using
the same procedure as above, including variations of the scales by the
factors $0.5$ and $2$ with respect to $\etg/2$.
 
\subsection{Theoretical uncertainties in the NNLO QCD prediction}
\begin{figure}[p]
\setlength{\unitlength}{1.0cm}
\begin{picture} (18.0,16.0)
\put (0.0,8.0){\includegraphics[width=8cm]{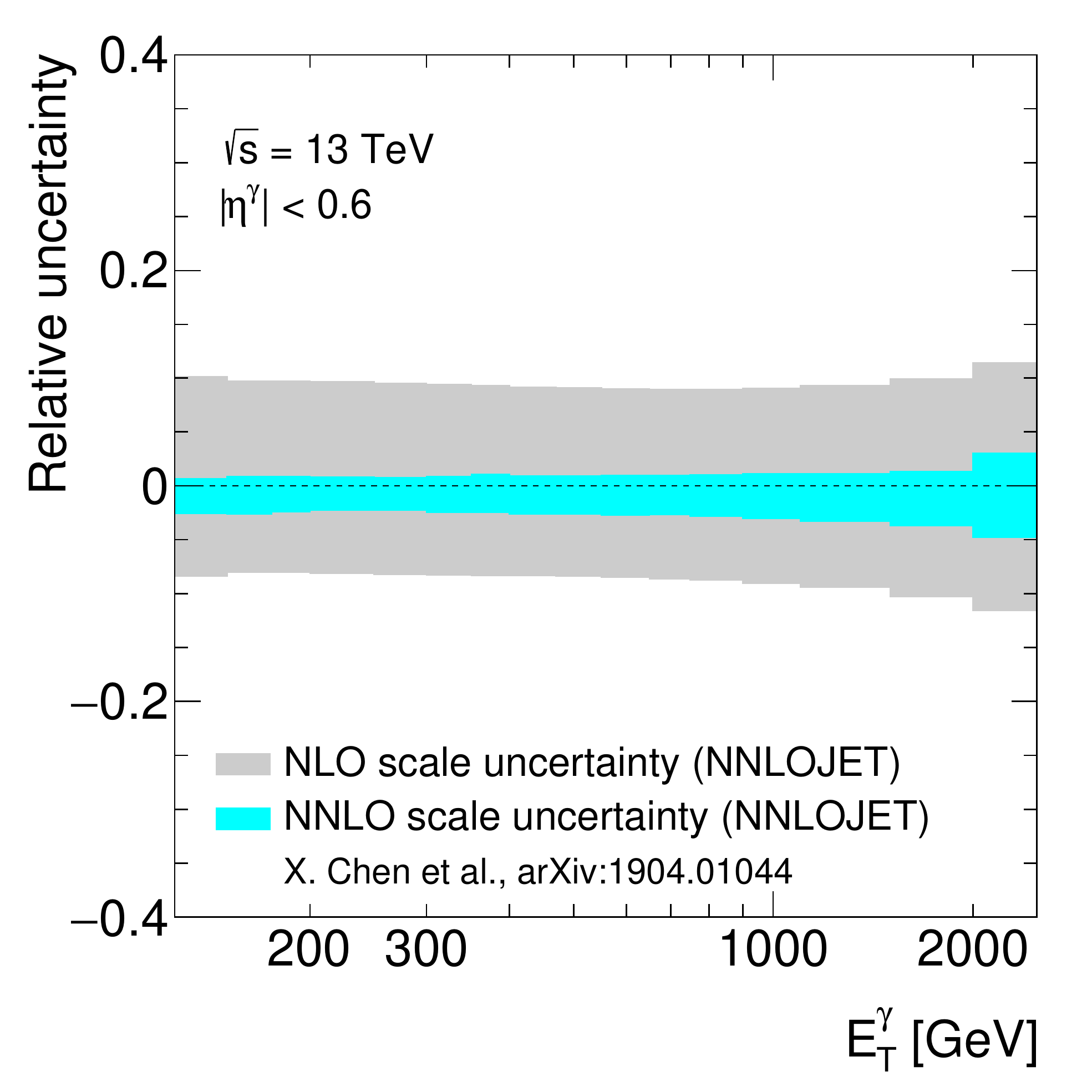}}
\put (8.0,8.0){\includegraphics[width=8cm]{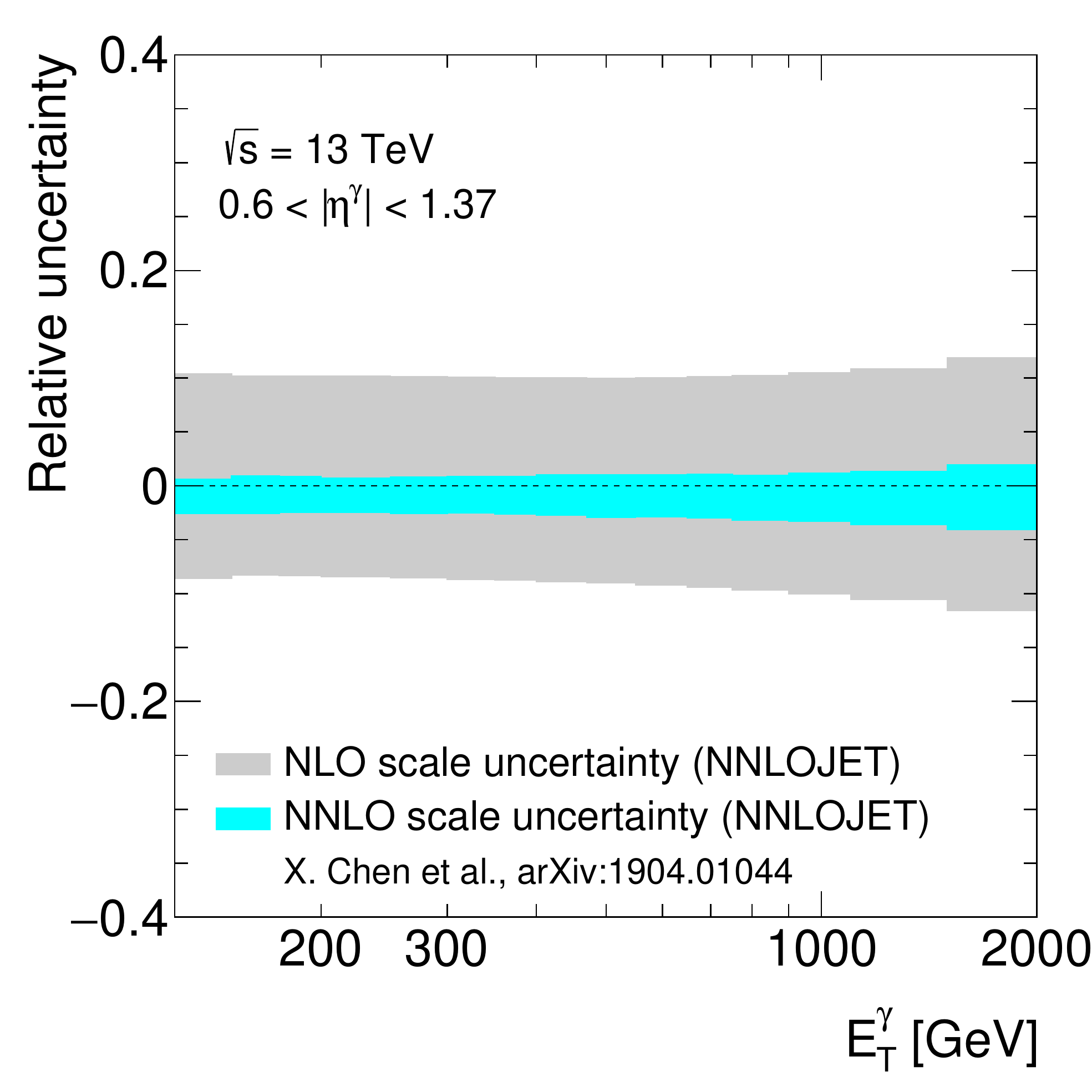}}
\put (0.0,0.0){\includegraphics[width=8cm]{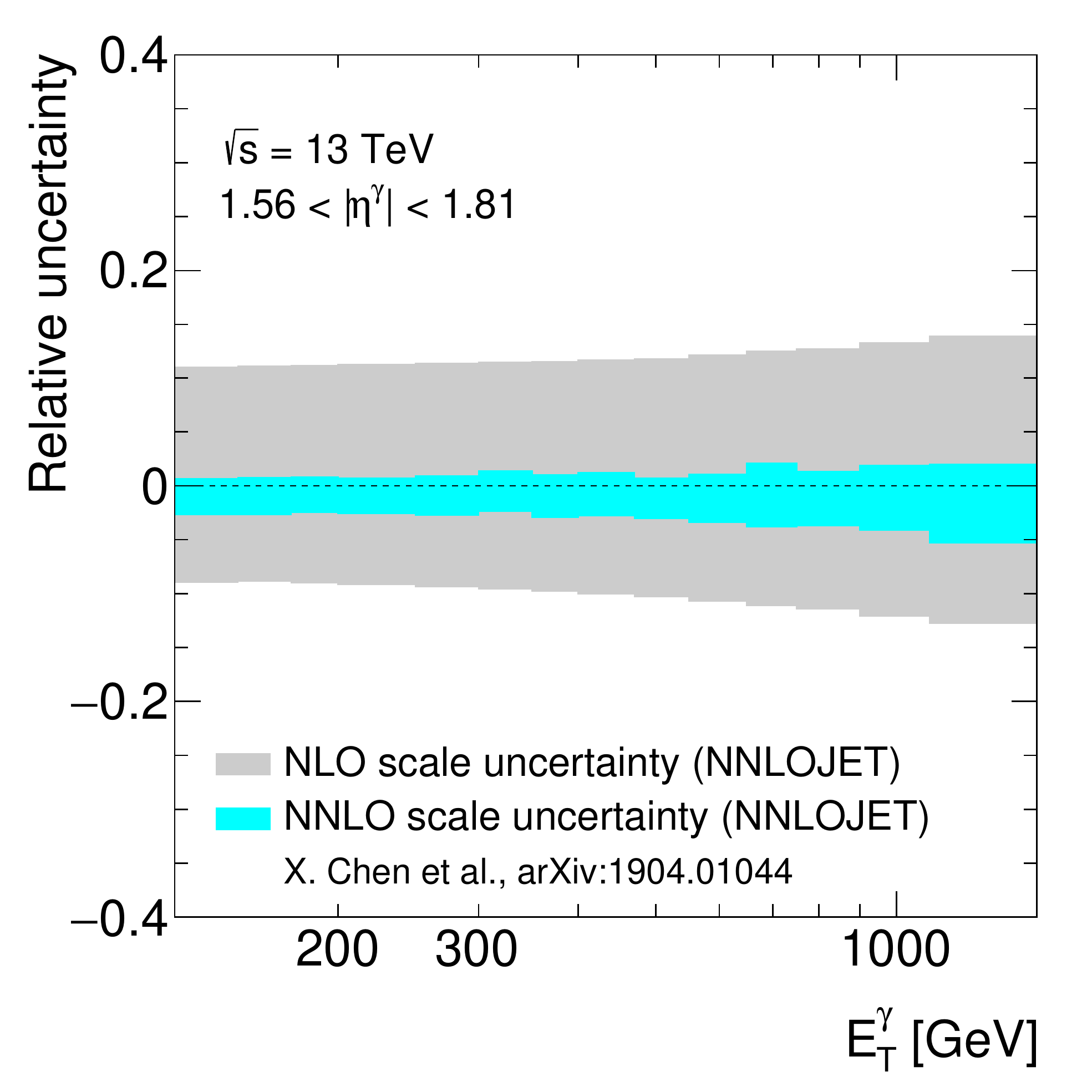}}
\put (8.0,0.0){\includegraphics[width=8cm]{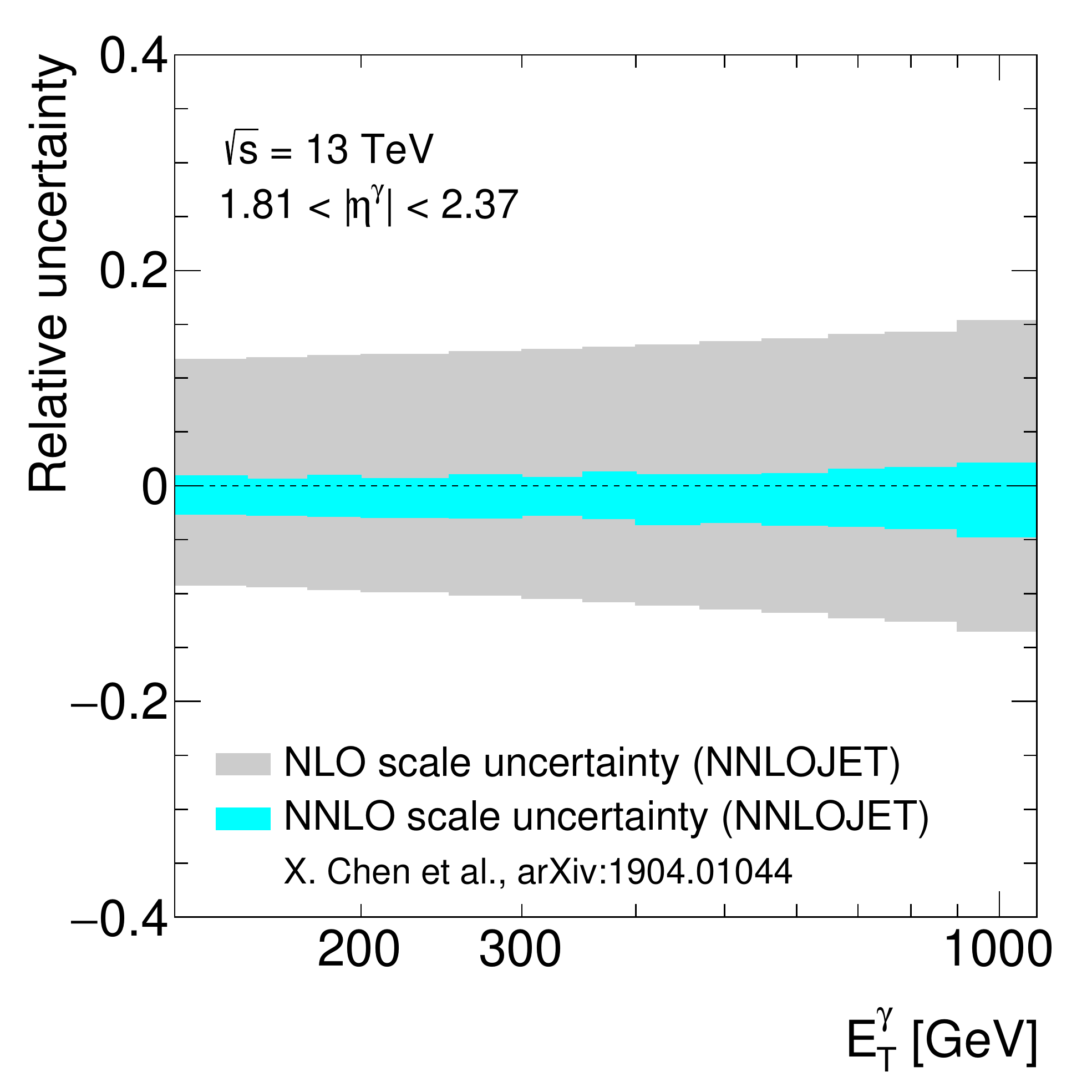}}
\end{picture}
\caption
{
The relative theoretical uncertainty in the cross-section prediction
of \nnlojet\ arising from the scale variations as a function of
$\etg$ in different regions of $|\etag|$: for NLO and NNLO QCD
predictions.
}
\label{fig5nnlo}
\end{figure}

The uncertainty in the NNLO QCD prediction due to higher-order terms
is evaluated by varying the renormalisation and factorisation scales
as is done for the \sher~2.2.2 prediction (see
Section~\ref{nlounc}). The resulting uncertainties are compared with
those of the NLO QCD prediction from \nnlojet\ in
Figure~\ref{fig5nnlo}. The uncertainties in the NNLO QCD prediction
arising from the scale variations are in the range $0.6\%$--$5\%$ and
are smaller than those in the NLO QCD prediction by a factor in the
range $2$--$20$, depending on $\etg$ and $\etag$. The uncertainties
due to the PDFs and $\as$ are not available, but they are expected to
be similar to, or smaller than, those affecting the \jetp\
prediction. For the comparison with the data in Section~\ref{res}, the
`total' uncertainty associated with the \nnlojet\ prediction is taken
as the sum in quadrature of the scale uncertainties, the statistical
uncertainties of the calculations, the uncertainty of $1\%$ assigned
to the non-perturbative corrections (see Section~\ref{nlounc}) and, as
an approximation, the uncertainties due to the PDFs and $\as$ as
estimated from \jetp\ at NLO.
 
\FloatBarrier
 
\section{Results}
\label{res}
\begin{figure}[h]
\setlength{\unitlength}{1.0cm}
\begin{picture} (18.0,14.5)
\put (0.0,-0.9){\centerline{\includegraphics[width=15cm]{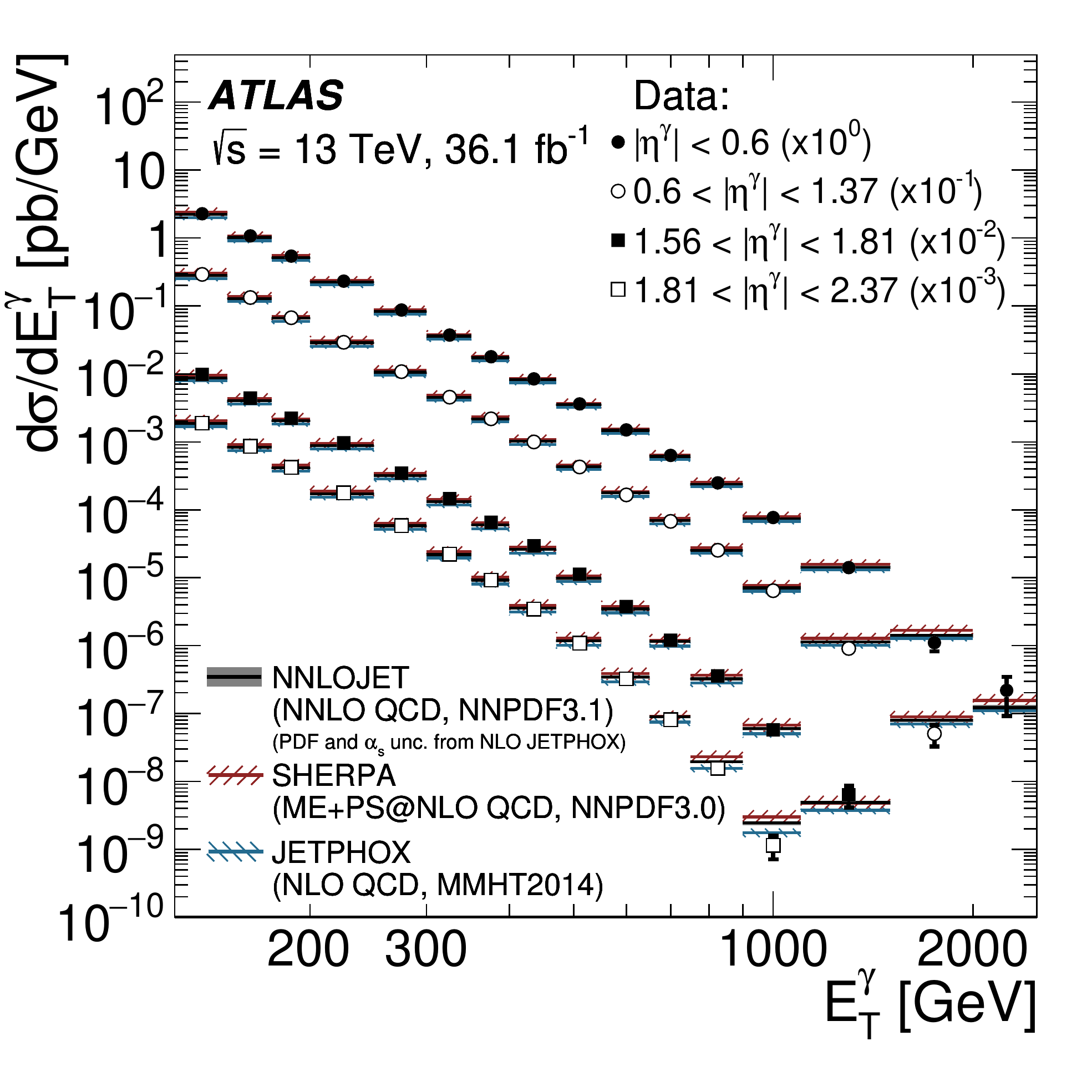}}}
\end{picture}
\caption
{
The measured differential cross section for isolated-photon
production as a function of $\etg$ in $|\etag|<0.6$ (black dots),
$0.6<|\etag|<1.37$ (open circles), $1.56<|\etag|<1.81$ (black
squares) and $1.81<|\etag|<2.37$ (open squares). The NLO pQCD
prediction from \jetp, the ME+PS@NLO QCD prediction from \sher~2.2.2
and the NNLO QCD prediction from \nnlojet\ are also shown. The
measurement and the predictions are normalised by the factors shown
in parentheses to aid visibility. The error bars represent the data
statistical uncertainties and systematic uncertainties added in
quadrature.  For most of the points, the error bars are smaller than
the marker size and, thus, not visible. The bands represent the
theoretical uncertainty associated with the predictions; in the case
of  \nnlojet, the uncertainties due to the PDFs and $\as$ are
estimated at NLO with \jetp.
}
\label{fig5}
\end{figure}

The measurement of the inclusive isolated-photon differential cross
section as a function of $\etg$ in different regions of $|\etag|$ is
shown in Figure~\ref{fig5}. In the region $|\etag|<0.6$
($0.6<|\etag|<1.37$) the cross section decreases by seven orders of
magnitude in the measured range $125 < \etg < 2500$~\GeV\ ($125 < \etg
< 2000$~\GeV). In the forward regions $1.56<|\etag|<1.81$ and
$1.81<|\etag|<2.37$, the measured ranges are $125 < \etg < 1500$~\GeV\
and $125 < \etg < 1100$~\GeV, respectively, and the cross section
spans approximately six orders of magnitude.
 
\begin{figure}[h]
\setlength{\unitlength}{1.0cm}
\begin{picture} (18.0,15.0)
\put (0.0,0.0){\centerline{\includegraphics[width=15cm]{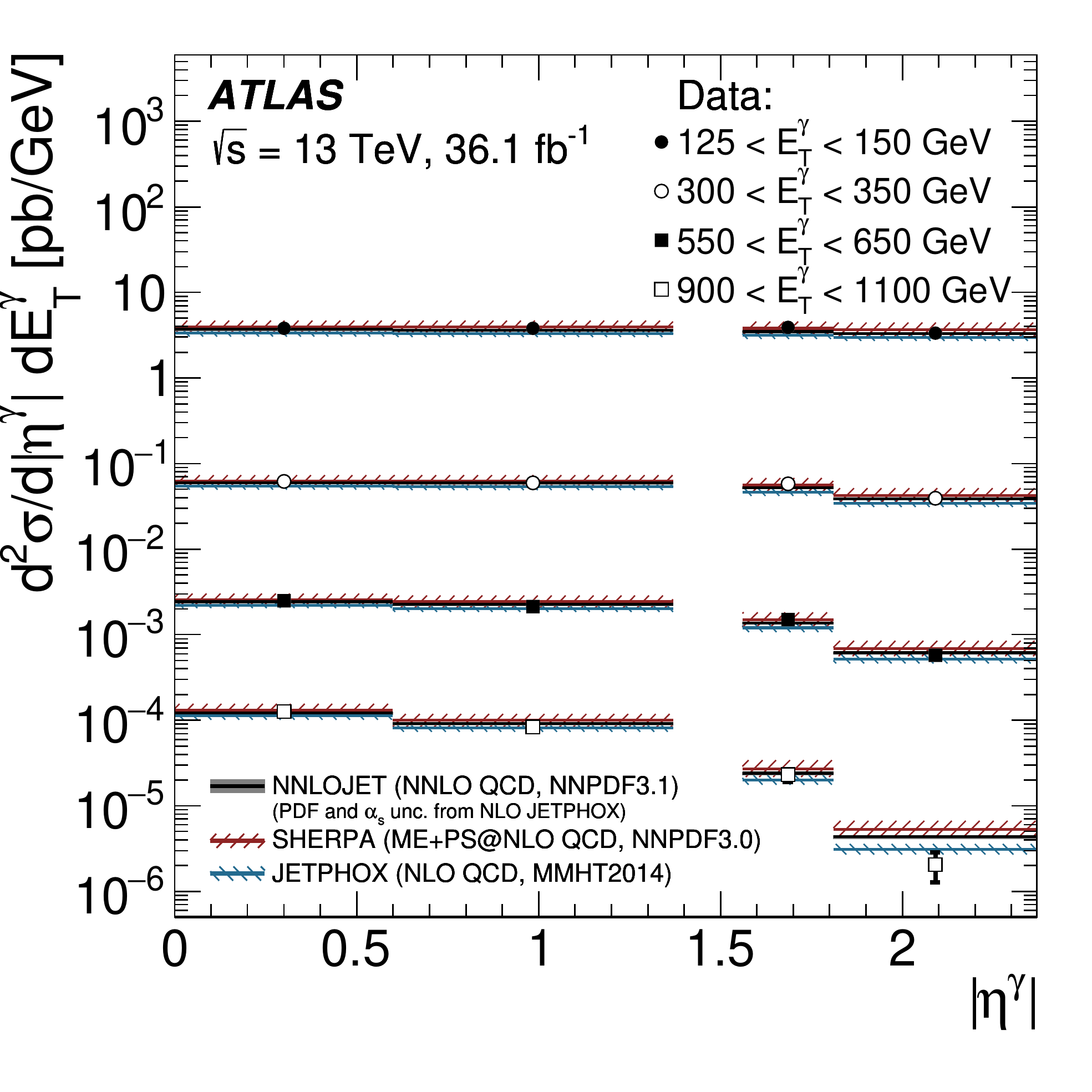}}}
\end{picture}
\caption
{
The measured double-differential cross section for isolated-photon
production as a function of $|\etag|$ in $125<\etg<150$~\GeV\ (black
dots), $300<\etg<350$~\GeV\ (open circles), $550<\etg<650$~\GeV\
(black squares) and $900<\etg<1100$~\GeV\ (open squares).  The NLO
pQCD prediction from \jetp, the ME+PS@NLO QCD prediction from
\sher~2.2.2 and the NNLO QCD prediction from \nnlojet\ are also
shown. The error bars represent the data statistical uncertainties
and systematic uncertainties added in quadrature.  For most of the
points, the error bars are smaller than the marker size and, thus,
not visible. The bands represent the theoretical uncertainty
associated with the predictions; in the case of  \nnlojet, the
uncertainties due to the PDFs and $\as$ are estimated at NLO with
\jetp.
}
\label{fig6}
\end{figure}

Figure~\ref{fig6} shows the inclusive isolated-photon
double-differential cross section
$\textrm{d}^2\sigma/\textrm{d}|\etag| \textrm{d}\etg$ as a function of
$|\etag|$ in different regions of $\etg$. At high $\etg$, the decrease
of the double-differential cross section with increasing $|\etag|$ is
more prominent. Whereas the measured double-differential cross section
decreases by $12\%$ from the most central point to the most forward
point for the lowest $\etg$ range, the decrease is $98\%$ for the
highest $\etg$ range.
 
The NLO QCD prediction of \jetp\  is compared with the measurement in
Figures~\ref{fig5} and \ref{fig6}. The NLO QCD prediction from
\sher~2.2.2 is also reported in Figures~\ref{fig5} and \ref{fig6} and
is denoted by `ME+PS@NLO QCD' to highlight the differences relative to
the predictions from \jetp\ discussed in Section~\ref{nlo}. The
nominal prediction of \sher~2.2.2 is above that of \jetp; this is
attributed to the fact that the former includes contributions from
parton showers, virtual corrections for $\gamma+ {\textrm{2-jet}}$ and
higher-order tree-level matrix elements for the processes
$2\rightarrow n$ with $n=4$ and $5$, which are not present in the
prediction of \jetp. Both types of predictions adequately reproduce
the dependence of the single- and double-differential cross section on
$\etg$ and $|\etag|$ as observed in the data.
 
\begin{figure}[h]
\setlength{\unitlength}{1.0cm}
\begin{picture} (18.0,17.0)
\put (0.0,0.0){\centerline{\includegraphics[width=17cm]{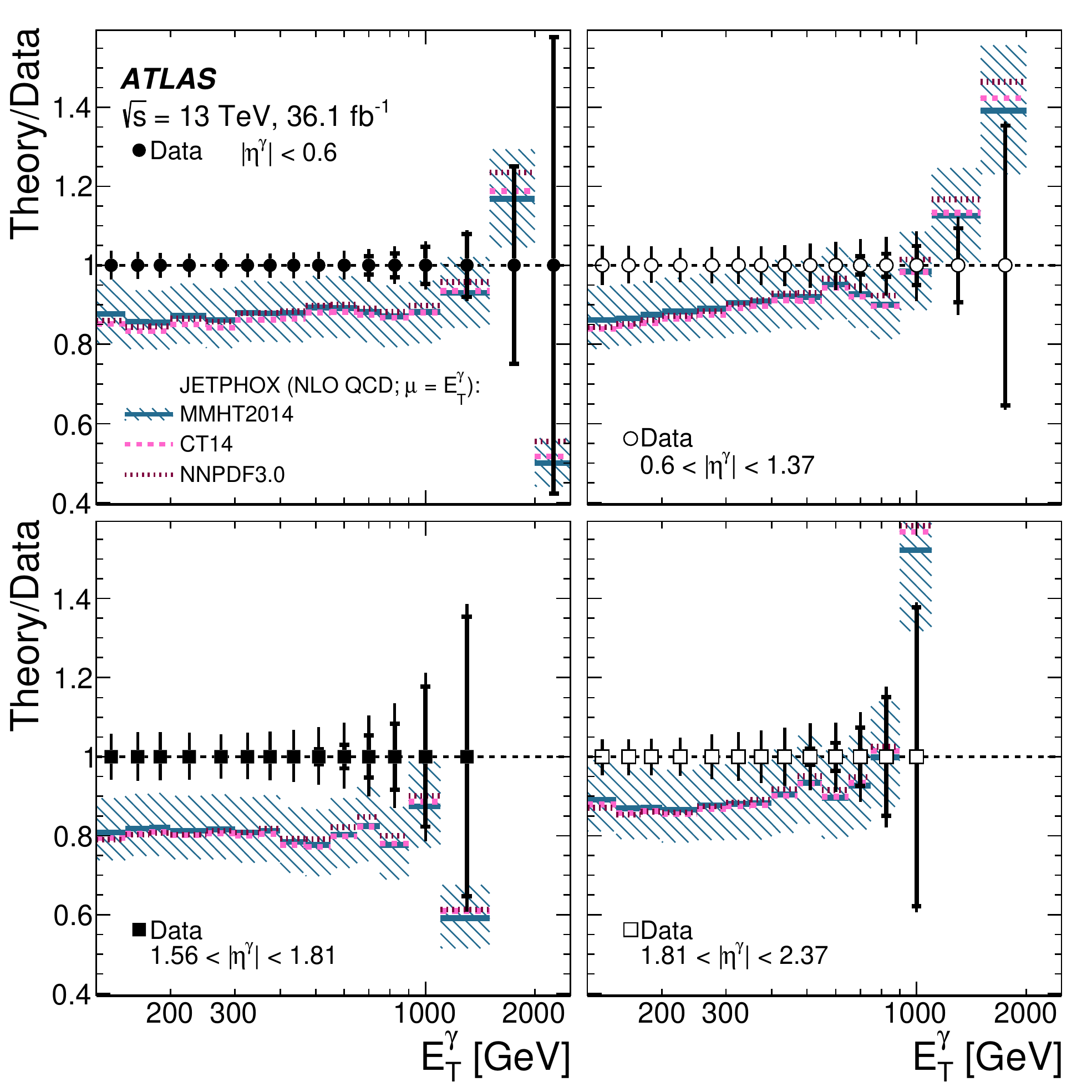}}}
\end{picture}
\caption
{
The ratio of the NLO pQCD prediction of {\jetp} with
$\muR=\muF=\muf=\etg$ using the MMHT2014 PDF set to the measured
differential cross section for isolated-photon production (solid
lines) as a function of $\etg$ in different regions of
$|\etag|$. The symbols for the data are centred at unity and the
inner (outer) error bars represent the relative data statistical
uncertainties (statistical and systematic uncertainties added in
quadrature). The hatched bands represent the full theoretical
uncertainty associated to the prediction based on the MMHT2014 PDF
set, including the variations of the scales. For comparison, the
predictions using the CT14 (dashed lines) and NNPDF3.0 (dotted
lines) PDF sets are also included.
}
\label{fig7}
\end{figure}

\begin{figure}[p]
\setlength{\unitlength}{1.0cm}
\begin{picture} (18.0,17.0)
\put (0.0,0.0){\centerline{\includegraphics[width=17cm]{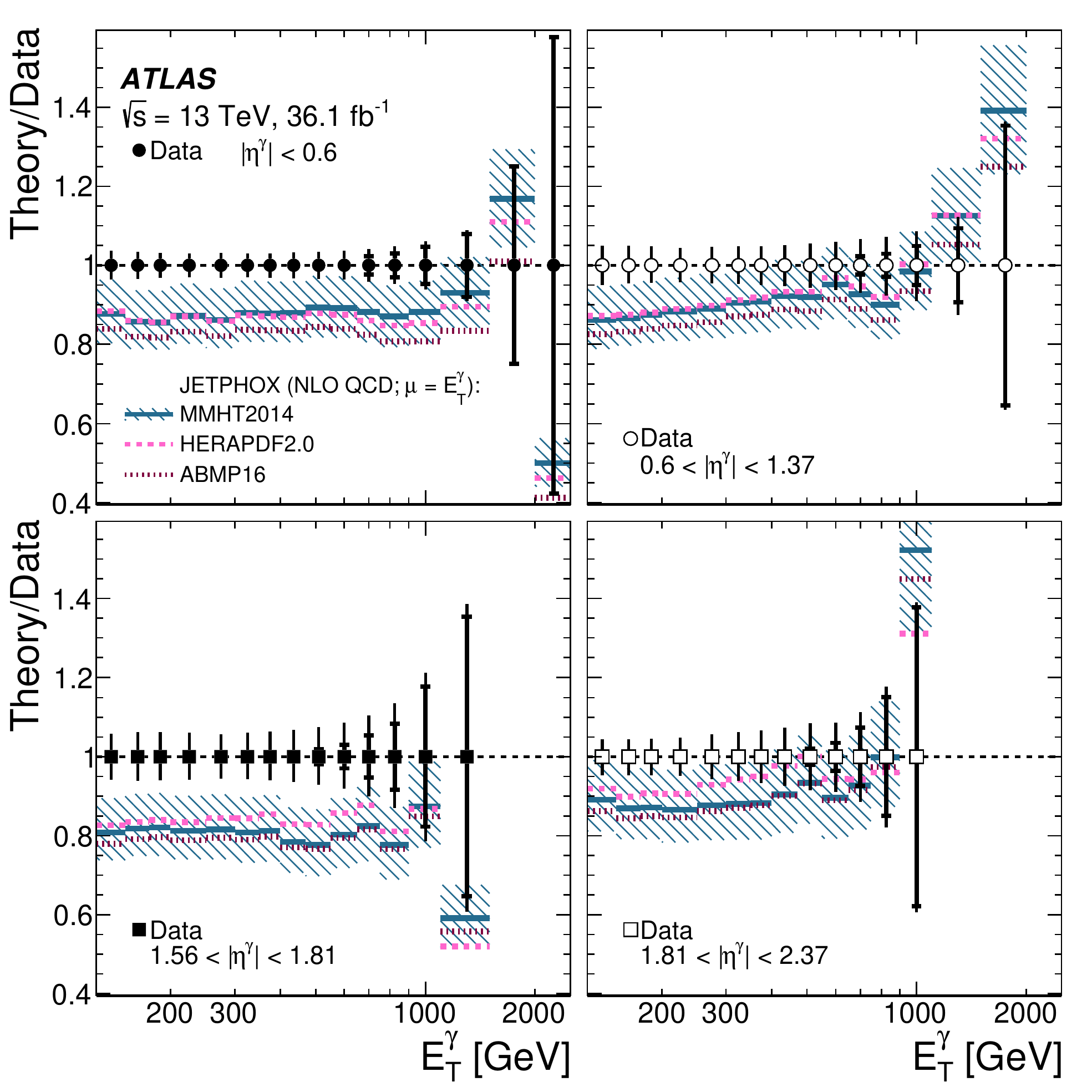}}}
\end{picture}
\caption
{
The ratio of the NLO pQCD prediction of {\jetp} with
$\muR=\muF=\muf=\etg$ using the MMHT2014 PDF set to the measured
differential cross section for isolated-photon production (solid
lines) as a function of $\etg$ in different regions of
$|\etag|$. The symbols for the data are centred at unity and the
inner (outer) error bars represent the relative data statistical
uncertainties (statistical and systematic uncertainties added in
quadrature). The hatched bands represent the full theoretical
uncertainty associated to the prediction based on the MMHT2014 PDF
set, including the variations of the scales. For comparison, the
predictions using the HERAPDF2.0 (dashed lines) and ABMP16 (dotted
lines) PDF sets are also included.
}
\label{fig8}
\end{figure}

\begin{figure}[p]
\setlength{\unitlength}{1.0cm}
\begin{picture} (18.0,17.0)
\put (0.0,0.0){\centerline{\includegraphics[width=17cm]{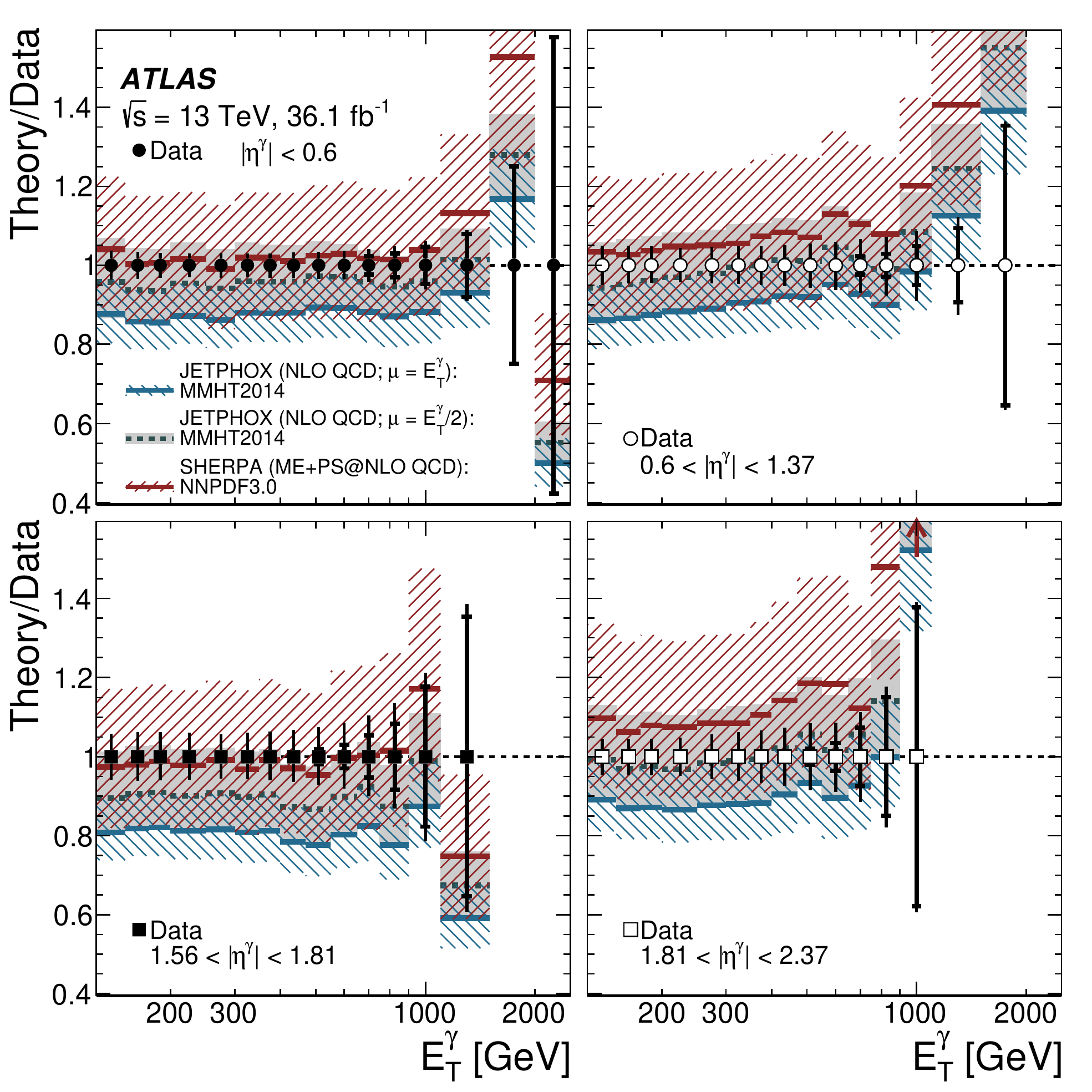}}}
\end{picture}
\caption
{
The ratio of the NLO pQCD prediction of {\jetp} with
$\muR=\muF=\muf=\etg$, the ratio of the NLO pQCD prediction of
{\jetp} with $\muR=\muF=\muf=\etg/2$ and the ratio of the ME+PS@NLO
QCD prediction of {\sher~2.2.2} to the measured differential cross
section for isolated-photon production as a function of $\etg$ in
different regions of $|\etag|$.  The symbols for the data are
centred at unity and the inner (outer) error bars represent the
relative data statistical uncertainties (statistical and systematic
uncertainties added in quadrature). The bands represent the full
theoretical uncertainty, which includes the variations of the
scales.
}
\label{fig17}
\end{figure}

Although the predictions are overall consistent with the measurement
within the uncertainties, a detailed comparison is presented in terms
of the ratios of theory predictions to data. The ratios of the
predictions of \jetp\ with $\muR=\muF=\muf=\etg$ using different PDF
sets to the measured cross section as a function of $\etg$ are shown
in Figures~\ref{fig7} and \ref{fig8}; the full theoretical
uncertainty, including the variations of the scales, is shown for the
prediction based on the MMHT2014 PDF set. The predictions based on the
MMHT2014, CT14 and NNPDF3.0 PDF sets are similar and closest to the
data for $|\etag|<1.37$ for most of the range in $\etg$. For
$1.56<|\etag|<2.37$, the prediction based on the HERAPDF2.0 PDF set is
closest to the data. The prediction based on the ABMP16 PDF set is
further away from the data than that based on the MMHT2014 PDF set for
$\etg \lesssim 1$~\TeV\ in the region $|\etag|<1.37$. The prediction
of \jetp\ with $\muR=\muF=\muf=\etg$ exhibits a tendency to
underestimate the data over most of the measured range in
$\etg$. However, the prediction of \jetp\ with a different choice for
the scales, namely $\muR=\muF=\muf=\etg/2$, provides an improved
description of the normalisation of the data, as can be seen in
Figure~\ref{fig17}; the full theoretical uncertainty, including the
variations of the scales with respect to $\etg/2$, is shown for the
prediction based on the MMHT2014 PDF set. The ratio of the prediction
of \sher~2.2.2 to the measured cross section as a function of $\etg$
is also shown in Figure~\ref{fig17}.  Good agreement is observed
between this prediction and the data distribution. In general, the NLO
pQCD predictions are consistent with the measurement within the
experimental and theoretical uncertainties.
 
\begin{figure}[p]
\setlength{\unitlength}{1.0cm}
\begin{picture} (18.0,17.0)
\put (0.0,0.0){\centerline{\includegraphics[width=17cm]{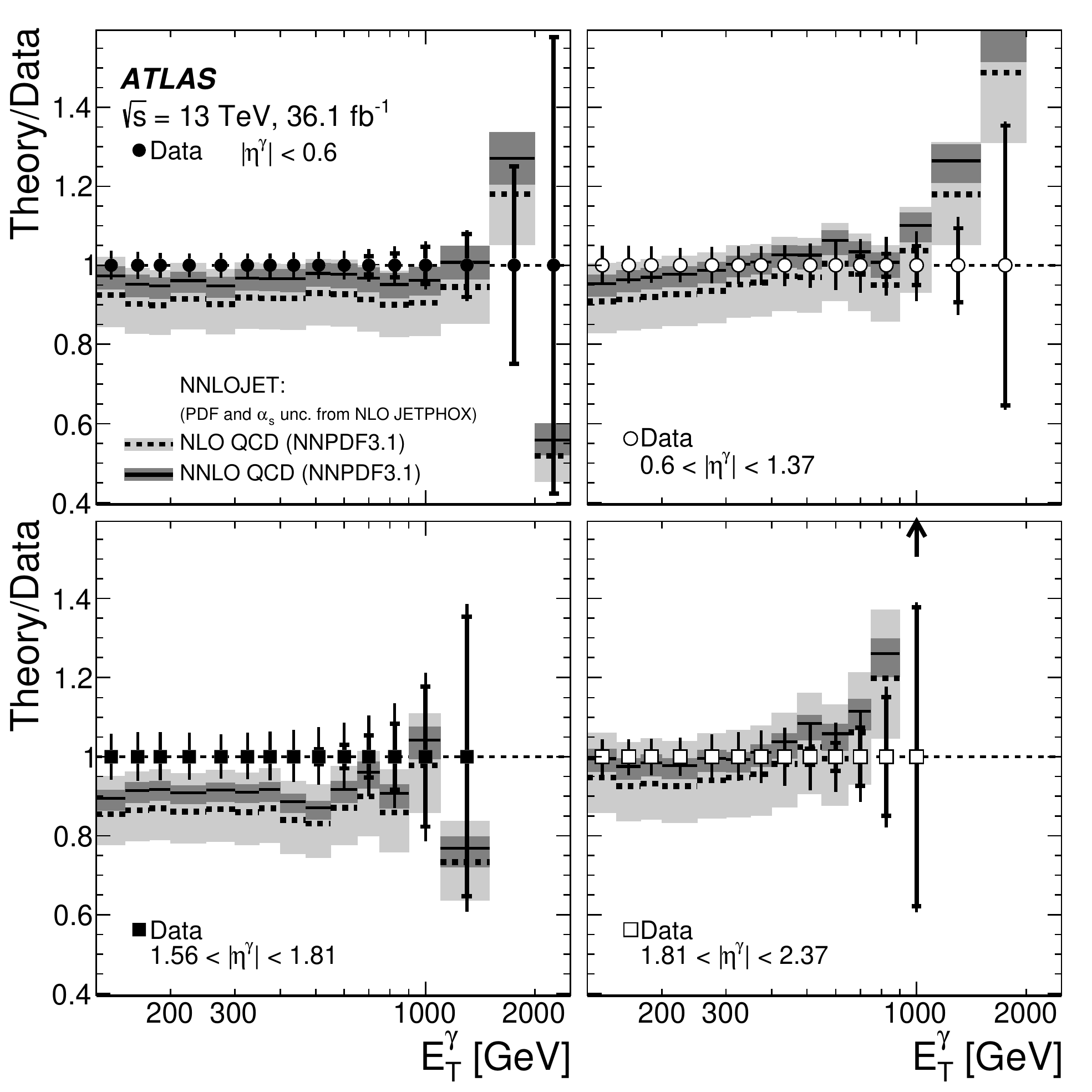}}}
\end{picture}
\caption
{
The ratio of the NNLO (NLO) QCD prediction of \nnlojet\ using the
NNPDF3.1 PDF set to the measured differential cross section for
isolated-photon production as a function of $\etg$ in different
regions of $|\etag|$ is shown as a solid (dashed) line.  The symbols
for the data are centred at unity and the inner (outer) error bars
represent the relative data statistical uncertainties (statistical
and systematic uncertainties added in quadrature). The shaded bands
represent the theoretical uncertainties, which include those  due to
the PDFs and $\as$ as estimated at NLO with \jetp.
}
\label{fig11}
\end{figure}

\begin{figure}[p]
\setlength{\unitlength}{1.0cm}
\begin{picture} (18.0,17.0)
\put (0.0,0.0){\centerline{\includegraphics[width=17cm]{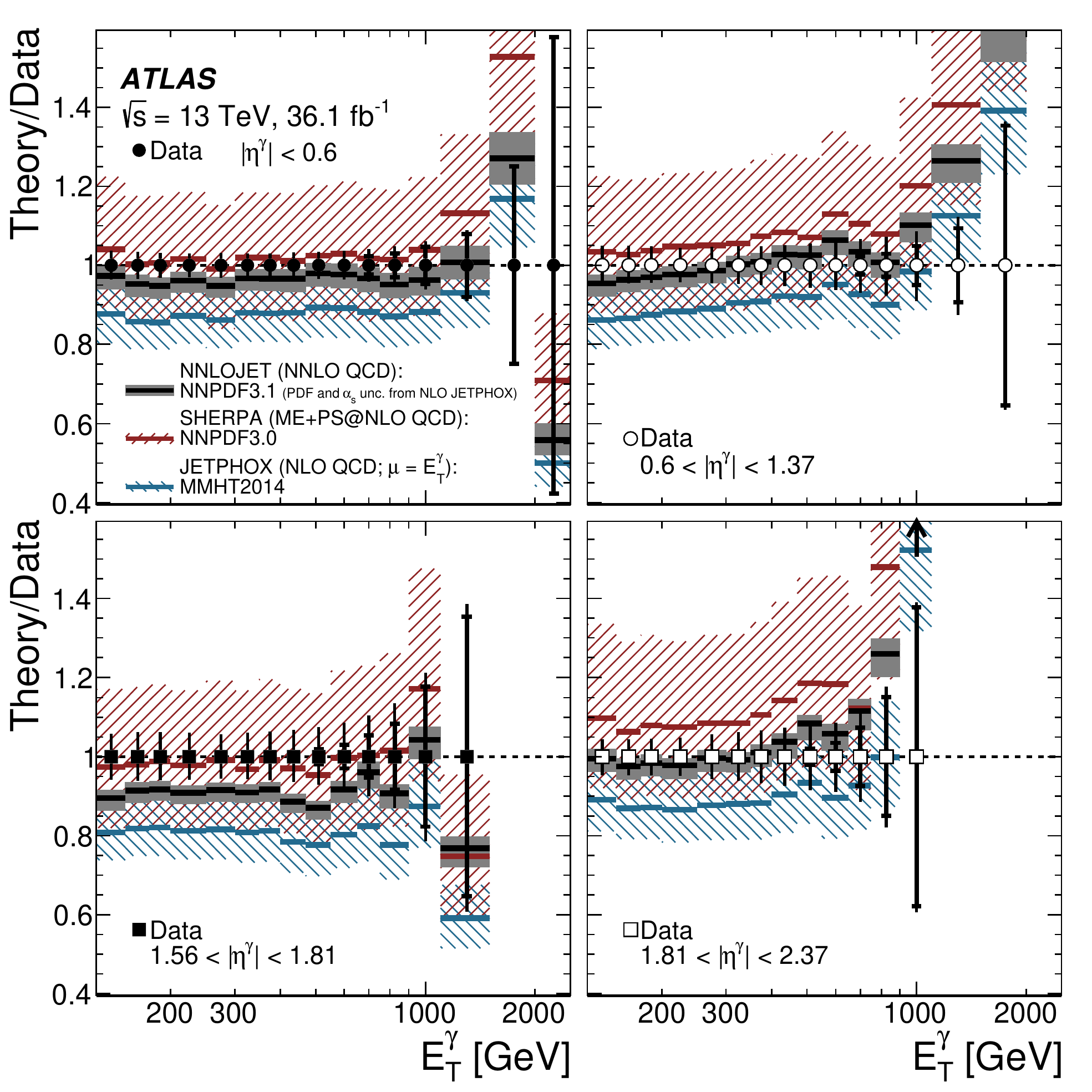}}}
\end{picture}
\caption
{
The ratio of the NLO pQCD prediction of {\jetp} with
$\muR=\muF=\muf=\etg$, the ratio of the ME+PS@NLO QCD prediction of
{\sher~2.2.2} and the ratio of the NNLO QCD prediction of \nnlojet\
to the measured differential cross section for isolated-photon
production as a function of $\etg$ in different regions of
$|\etag|$.  The symbols for the data are centred at unity and the
inner (outer) error bars represent the relative data statistical
uncertainties (statistical and systematic uncertainties added in
quadrature). The bands represent the theoretical uncertainty; in the
case of \nnlojet, the uncertainties due to the PDFs and $\as$ are
estimated at NLO with \jetp.
}
\label{fig18}
\end{figure}

The NNLO QCD prediction from \nnlojet, which represent a significant
step forward in precision, is compared with the data in
Figures~\ref{fig5}, \ref{fig6}, \ref{fig11} and \ref{fig18}. The
inclusion of the NNLO QCD corrections increases the normalisation of
the prediction and significantly reduces the uncertainties due to the
scale variations, which are dominant for NLO QCD calculations. With
the described choice of the input parameters, the NNLO QCD prediction
provides an excellent description of the data except in the region
$1.56 < |\etag|<1.81$, where there is a tendency in the theory to
underestimate the data. The improvements in the description of the
data by the NNLO QCD prediction relative to that of NLO QCD are shown
in Figures~\ref{fig11} and \ref{fig18}.
 
\FloatBarrier
 
\section{Summary and conclusions}
\label{conc}
The cross section for inclusive isolated-photon production in $pp$
collisions at $\sqrt s=13$~\TeV\ is measured using data with an
integrated luminosity of $36.1$~\fb1\ collected by the ATLAS detector
at the LHC. The differential cross section as a function of $\etg$ is
measured in different regions of $|\etag|$ for photons with
$\etg>125$~\GeV\ and $|\etag|<2.37$, excluding the region
$1.37<|\etag|<1.56$.
 
The measurement presented here constitutes an improvement in terms of
statistical and systematic uncertainties relative to that published
earlier thanks to the ten-fold increase in the integrated
luminosity. In the new measurement,  the reach in $\etg$ is extended
upwards by $1000,\ 500,\ 400$ and $200$~\GeV\ for $|\etag|<0.6$,
$0.6<|\etag|<1.37$, $1.56<|\etag|<1.81$ and $1.81<|\etag|<2.37$,
respectively, relative to the previous measurement at $\sqrt
s=13$~\TeV. The range in which the measurement is limited by
systematic uncertainties is extended upwards to $1$~\TeV\ for
$|\etag|<1.37$, where previously the limitation was at $600$~\GeV.
 
Next-to-leading-order QCD predictions using several PDF sets are
compared with the measurement and found to provide an adequate
description of the data within the experimental and theoretical
uncertainties. The comparison of data and theory is limited by the
theoretical uncertainties due to terms beyond NLO in QCD.
Experimental systematic uncertainties are smaller than the theoretical
uncertainties over the full investigated phase space.  The inclusion
of higher-order matrix elements and parton showers, as provided by the
\sher~2.2.2 program, improves the description of the normalisation of
the data by the prediction.
 
Recently, a next-to-next-to-leading-order QCD prediction was
calculated in which the uncertainties due to the scale variations are
reduced by a factor in the range $2$--$20$ relative to those of a NLO
QCD calculation. Overall, the NNLO QCD prediction gives an excellent
description of the data. The comparison of the NNLO QCD prediction
with the measured cross section represents a precise test of the
theory at ${\cal O}(\alpha_{\textsc{em}}\alpha^3_{\mathrm{s}})$ in the
range of photon transverse energies from $125$~\GeV\ up to and beyond
$1$~\TeV. The measurement has the potential to further constrain the
PDFs, particularly the gluon density in the proton, within a global
NNLO QCD fit.
 
\section*{Acknowledgements}
 
 
We thank CERN for the very successful operation of the LHC, as well as the
support staff from our institutions without whom ATLAS could not be
operated efficiently.
 
We acknowledge the support of ANPCyT, Argentina; YerPhI, Armenia; ARC, Australia; BMWFW and FWF, Austria; ANAS, Azerbaijan; SSTC, Belarus; CNPq and FAPESP, Brazil; NSERC, NRC and CFI, Canada; CERN; CONICYT, Chile; CAS, MOST and NSFC, China; COLCIENCIAS, Colombia; MSMT CR, MPO CR and VSC CR, Czech Republic; DNRF and DNSRC, Denmark; IN2P3-CNRS, CEA-DRF/IRFU, France; SRNSFG, Georgia; BMBF, HGF, and MPG, Germany; GSRT, Greece; RGC, Hong Kong SAR, China; ISF and Benoziyo Center, Israel; INFN, Italy; MEXT and JSPS, Japan; CNRST, Morocco; NWO, Netherlands; RCN, Norway; MNiSW and NCN, Poland; FCT, Portugal; MNE/IFA, Romania; MES of Russia and NRC KI, Russian Federation; JINR; MESTD, Serbia; MSSR, Slovakia; ARRS and MIZ\v{S}, Slovenia; DST/NRF, South Africa; MINECO, Spain; SRC and Wallenberg Foundation, Sweden; SERI, SNSF and Cantons of Bern and Geneva, Switzerland; MOST, Taiwan; TAEK, Turkey; STFC, United Kingdom; DOE and NSF, United States of America. In addition, individual groups and members have received support from BCKDF, CANARIE, CRC and Compute Canada, Canada; COST, ERC, ERDF, Horizon 2020, and Marie Sk{\l}odowska-Curie Actions, European Union; Investissements d' Avenir Labex and Idex, ANR, France; DFG and AvH Foundation, Germany; Herakleitos, Thales and Aristeia programmes co-financed by EU-ESF and the Greek NSRF, Greece; BSF-NSF and GIF, Israel; CERCA Programme Generalitat de Catalunya, Spain; The Royal Society and Leverhulme Trust, United Kingdom.
 
The crucial computing support from all WLCG partners is acknowledged gratefully, in particular from CERN, the ATLAS Tier-1 facilities at TRIUMF (Canada), NDGF (Denmark, Norway, Sweden), CC-IN2P3 (France), KIT/GridKA (Germany), INFN-CNAF (Italy), NL-T1 (Netherlands), PIC (Spain), ASGC (Taiwan), RAL (UK) and BNL (USA), the Tier-2 facilities worldwide and large non-WLCG resource providers. Major contributors of computing resources are listed in Ref.~\cite{ATL-GEN-PUB-2016-002}.
 
 
\printbibliography
 
\newpage
\input{atlas_authlist} 
\end{document}

%% file: atlas_authlist.tex
 
\begin{flushleft}
{\Large The ATLAS Collaboration}

\bigskip

G.~Aad$^\textrm{\scriptsize 101}$,    
B.~Abbott$^\textrm{\scriptsize 128}$,    
D.C.~Abbott$^\textrm{\scriptsize 102}$,    
O.~Abdinov$^\textrm{\scriptsize 13,*}$,    
A.~Abed~Abud$^\textrm{\scriptsize 70a,70b}$,    
K.~Abeling$^\textrm{\scriptsize 53}$,    
D.K.~Abhayasinghe$^\textrm{\scriptsize 93}$,    
S.H.~Abidi$^\textrm{\scriptsize 167}$,    
O.S.~AbouZeid$^\textrm{\scriptsize 40}$,    
N.L.~Abraham$^\textrm{\scriptsize 156}$
% ...
\bigskip
\\

% ...
$^{13}$Institute of Physics, Azerbaijan Academy of Sciences, Baku; Azerbaijan.\\
% ...
$^{40}$Niels Bohr Institute, University of Copenhagen, Copenhagen; Denmark.\\
% ...
$^{53}$II. Physikalisches Institut, Georg-August-Universit\"{a}t G\"ottingen, G\"ottingen; Germany.\\
% ...
$^{70}$$^{(a)}$INFN Sezione di Pavia;$^{(b)}$Dipartimento di Fisica, Universit\`a di Pavia, Pavia; Italy.\\
% ...
$^{93}$Department of Physics, Royal Holloway University of London, Egham; United Kingdom.\\
% ...
$^{101}$CPPM, Aix-Marseille Universit\'e, CNRS/IN2P3, Marseille; France.\\
$^{102}$Department of Physics, University of Massachusetts, Amherst MA; United States of America.\\
% ...
$^{128}$Homer L. Dodge Department of Physics and Astronomy, University of Oklahoma, Norman OK; United States of America.\\
% ...
$^{156}$Department of Physics and Astronomy, University of Sussex, Brighton; United Kingdom.\\
% ...
$^{167}$Department of Physics, University of Toronto, Toronto ON; Canada.\\
% ...
$^{a}$ Also at Borough of Manhattan Community College, City University of New York, New York NY; United States of America.\\
$^{b}$ Also at Centre for High Performance Computing, CSIR Campus, Rosebank, Cape Town; South Africa.\\
% ...
$^{*}$ Deceased

\end{flushleft}
